\documentclass[useAMS,usenatbib]{mn2e}

\usepackage{graphicx}
\usepackage{mathrsfs}
\usepackage{natbib}
\usepackage{amssymb}
\usepackage{amsmath}
\usepackage{mathtools}
\usepackage{overpic}
\usepackage{color}
\usepackage[version=3]{mhchem}
\usepackage[normalem]{ulem} 
\usepackage{booktabs}
\usepackage{tabularx}
\usepackage{natbib}
\usepackage{comment}
\usepackage[breaklinks=true]{hyperref}
\usepackage[noabbrev]{cleveref}

\def\sun{{\odot}}
\newcommand{\skynet}{\emph{SkyNet}}
\newcommand{\smce}[1]{\smash{\ce{#1}}}

\newcommand{\newacronym}[3]{\newcommand{#1}[1]{#3##1 (#2##1)%
\renewcommand{#1}[1]{#2####1}}}

\newacronym{\NSE}{NSE}{nuclear statistical equilibrium}
\newacronym{\EOS}{EOS}{equation of state}
\newacronym{\sGRB}{sGRB}{short gamma ray burst}
\newacronym{\HST}{HST}{Hubble Space Telescope}
\newacronym{\HMNS}{HMNS}{hypermassive neutron star}
\newacronym{\BH}{BH}{black hole}
\newacronym{\ISCO}{ISCO}{innermost stable circular orbit}

\newcommand{\ye}{Y_{e,\text{5GK}}}

\crefname{equation}{Equation}{Equations}
\crefname{figure}{Figure}{Figures}
\crefname{table}{Table}{Tables}

\begin{document}


\title[Nucleosynthesis in HMNS disk winds]{Signatures of hypermassive neutron
star lifetimes on r-process nucleosynthesis in the disk ejecta from neutron
star mergers}
\author[Lippuner et al.]{Jonas Lippuner$^1$, Rodrigo Fern\'andez$^{2,3,4}$,
 Luke F. Roberts$^{5,6}$, Francois Foucart$^{7,8}$,
 \and
 Daniel Kasen$^{3,4,8}$, Brian D.\ Metzger$^9$, Christian D.\ Ott$^{1,10}$ \\
 $^1$ TAPIR, Walter Burke Institute for Theoretical Physics, California
 Institute of Technology, Pasadena, CA 91125, USA\\
 $^2$ Department of Physics, University of Alberta, Edmonton, AB T6G 2E1,
 Canada \\
 $^3$ Department of Physics, University of California, Berkeley, CA 94720,
 USA \\
 $^4$ Department of Astronomy \& Theoretical Astrophysics Center, University of
 California, Berkeley, CA 94720, USA \\
 $^5$ National Superconducting Cyclotron Laboratory, Michigan State University,
 East Lansing, MI 48824, USA \\
 $^6$ Department of Physics and Astronomy, Michigan State University, East
 Lansing, MI 48824, USA \\
 $^7$ NASA Einstein Fellow \\
 $^8$ Nuclear Science Division, Lawrence Berkeley National Laboratory, Berkeley,
 CA 94720, USA \\
 $^9$ Columbia Astrophysics Laboratory, Columbia University, New York, NY 10027,
 USA \\
 $^{10}$ Center for Gravitational Physics and International Research Unit of Advanced Future Studies, \\ \hspace*{1em} Yukawa Institute for Theoretical Physics, Kyoto University, Kyoto,  Kyoto Prefecture 606-8317, Japan
 }

 \date{Submitted to MNRAS}
 \pagerange{1--18}
 \pubyear{2017}
 \maketitle


\begin{abstract}
We investigate the nucleosynthesis of heavy elements in the winds ejected by
accretion disks formed in neutron star mergers. We compute
the element formation in disk outflows from \HMNS{} remnants of variable
lifetime, including the effect of angular momentum transport in the disk
evolution. We employ long-term axisymmetric hydrodynamic disk simulations to
model the ejecta, and compute r-process nucleosynthesis with tracer particles using a
nuclear reaction network containing $\sim 8000$ species. We find that the
previously known strong correlation between \HMNS{} lifetime, ejected mass,
and average electron fraction in the outflow is directly related to the amount
of neutrino irradiation on the disk, which dominates mass ejection at early
times in the form of a neutrino-driven wind. Production of lanthanides and
actinides saturates at short \HMNS{} lifetimes ($\lesssim 10$~ms), with
additional ejecta contributing to a blue optical kilonova component for
longer-lived \HMNS{s}. We find good agreement between the abundances from the disk
outflow alone and the solar r-process distribution only for short
\HMNS{} lifetimes ($\lesssim 10$~ms). For longer lifetimes, the rare-earth and
third r-process peaks are significantly under-produced compared to the solar
pattern, requiring additional contributions from the dynamical ejecta. The
nucleosynthesis signature from a spinning \BH{} can only overlap with
that from a \HMNS{} of moderate lifetime ($\lesssim 60$~ms). Finally, we show
that angular momentum transport not only contributes with a late-time outflow
component, but that it also enhances the neutrino-driven component by moving
material to shallower regions of the gravitational potential, in addition to
providing additional heating.
\end{abstract}


\maketitle

\section{Introduction}

The astrophysical origin of chemical elements formed through the rapid neutron
capture process (r-process) remains an open problem in nuclear astrophysics.
Observed abundances in metal-poor Galactic halo stars demand a mechanism that
produced a robust abundance pattern -- mirroring that in the Solar System -- for
elements with mass number $A\gtrsim 130$ \citep[e.g.,][]{sneden:08}. This
mechanism must also have operated since early on in cosmic history, since
r-process elements are found in very old metal-poor stars
\citep[e.g.,][]{cowan:99,ji:16}. In contrast, observed abundances of light
r-process elements in metal-poor stars show deviations from the solar system pattern relative to
heavier elements \citep[e.g.,][]{montes:07}. Meteoritic abundances point to
different formation timescales for light and heavy r-process elements,
suggesting that there might be more than one dominant formation site
\citep{wasserburg:96}.

Core-collapse supernovae may be able to produce light r-process elements
($A\lesssim 130$), but recent work indicates that they are most likely not the
dominant source of heavy r-process elements \citep[e.g.,][]{roberts:10,%
fischer:10,huedepohl:10,martinezpinedo:12,wanajo:13}. On the other hand,
mergers of binaries containing two neutron stars (NSNS) or a neutron star and a
black hole (NSBH) have long been considered as candidate r-process sites,
given the highly neutron-rich conditions achieved in the ejected material
\citep{lattimer:74}. The study of NSNS/NSBH mergers has intensified in recent
years given that (1) they are likely to be detected in gravitational waves by
Advanced LIGO/Virgo within the next few years
\citep[e.g.,][]{abadie:10,aligo:15}, (2) recent developments in numerical
relativity have enabled merger simulations consistent with Einstein's equations
of general relativity \citep[e.g.,][]{lehner:14,paschalidis:16}, and (3) the
electromagnetic signal from these events can aid with the localization of these
sources and provide information complementary to that carried by gravitational
waves \citep[e.g.,][]{rosswog:15b,fernandez:16,tanaka:16}.

Recent work has shown that the dynamical ejecta from NSNS/NSBH mergers can
produce a robust Solar abundance pattern for $A\gtrsim 130$ by virtue of
\emph{fission cycles} \citep[e.g.,][]{goriely:05}, with little sensitivity to
binary parameters or the equation of state
\citep[e.g.,][]{goriely:11,korobkin:12b,bauswein:13}, depending instead on
nuclear physics properties such as the fission fragment distribution
\citep[e.g.,][]{eichler:15}. When neutrino absorption is included in the
calculations, a larger fraction of lighter elements ($A < 130$) can be obtained
\citep[e.g.,][]{wanajo:14,goriely:15,sekiguchi:15,radice:16b,foucart:16,%
foucart:16a,roberts:16b}.

In addition to the dynamical ejecta, the accretion disk formed in NSNS/NSBH
mergers can eject a significant amount of material on timescales longer than the
dynamical time \citep[e.g.,][]{ruffert:97,lee:09b,metzger:09b}. The
neutron-to-seed ratio of this material is lower than that in the dynamical
ejecta, because the longer evolutionary timescales allow weak interactions to modify the
composition more significantly \citep[e.g.,][]{dessart:09,fernandez:13,perego:14}. Matter
can be ejected on the thermal timescale of the disk ($\sim 30$~ms,
\S\ref{s:investigated_models}) by neutrino energy deposition, or on much longer
timescales ($\sim 1$~s, \S\ref{s:investigated_models}) by a combination of
angular momentum transport processes and nuclear recombination. The amount of
mass ejected by the disk on the longer timescale can be comparable to that in the dynamical
ejecta \cite[e.g.,][]{fernandez:16}, while the neutrino-driven component is
significant only when a \HMNS{} phase precedes \BH{} formation
\citep{dessart:09,fernandez:13,metzger:14,just:15}.

Early work on nucleosynthesis from NSNS/NSBH merger disks focused on
neutrino-driven outflows and used a parametric treatment to obtain thermodynamic
trajectories for composition analysis \citep[e.g.,][]{surman:08,wanajo:11},
finding that conditions for both light and heavy r-process elements are possible
in these outflows. More recent work has employed tracer particles from long-term
hydrodynamic simulations of the disk when a \BH{} is the central object
\citep{just:15,wu:16}. These studies have found that the outflow generates a
robust abundance of elements around $A=130$, with significant production of
lighter elements, and a variable yield of heavy r-process elements that depends
on binary properties and disk physics. The case of a \HMNS{} at the center was
studied by \cite{martin:15} using tracer particles from a time-dependent
simulation that considered the neutrino-driven wind phase \citep{perego:14}. The
resulting outflow generates primarily nuclei with $A < 130$, with a significant
dependence of the yield on latitude and ejection time. The long-term (viscous)
outflow was not captured, however, and the lifetime of the \HMNS{} was accounted
for only by looking at subsets of particles that were ejected before a certain
time.

In this study, we analyze the nucleosynthesis yields from the long-term outflow
generated by an accretion disk around a \HMNS{} of variable lifetime. Following
the approach of \cite{metzger:14}, we conduct a number of long-term disk
simulations in which the \HMNS{} transforms into a \BH{} at different times.
This parameterized approach does not require the assumption of a particular
equation of state of dense matter. It is also independent of the complex processes that determine transport of
angular momentum and cooling, all of which set the survival time of the \HMNS{}
\citep[e.g.,][]{paschalidis:12,kaplan:14}. In order to obtain nucleosynthesis yields,
tracer particles are injected in the disk initially, and the resulting
thermodynamic trajectories are analyzed with the nuclear reaction network
\skynet\ \citep{lippuner:17b}.

This paper is structured as follows. Section~\ref{s:methods} describes the
numerical method employed and the models studied. Results are presented and
discussed in section~\ref{s:results}, and we conclude in
section~\ref{s:conclusions}. The appendix contains further details about our
numerical implementation.

\section{Methods}
\label{s:methods}

\subsection{Disk outflow simulations and thermodynamic trajectories}
\label{s:flash_description}

The long-term evolution of the accretion disk is computed using
the approach described in \citet{metzger:14}. In short, the equations of Newtonian
hydrodynamics and lepton number conservation are solved with the FLASH code
\citep{fryxell:00,dubey:09} assuming axisymmetry (2D). Source terms include
gravity via a pseudo-Newtonian potential, shear viscosity with an $\alpha$
prescription \citep{shakura:73}, and contributions due to neutrino absorption
and emission to the lepton number and energy equations. A pseudo-Newtonian
potential, such as that from \citet{artemova:96}, approximates the effect of a
Schwarzschild or Kerr metric by providing an \ISCO{} in an otherwise Newtonian
hydrodynamic simulation. Other aspects, such as the angular dependence of
the Kerr metric, are not captured.

Neutrinos are implemented in a leakage scheme for cooling and a disk light-bulb
approximation for self-irradiation. Only charged-current interactions are
included, since they have the largest cross-section, exchange energy with
matter, and drive the evolution of the electron fraction. The \HMNS{} is
approximated as a reflecting sphere with a radius of $30$~km and a rotation
period of $1.5$~ms. A time-dependent, isotropic outward neutrino flux is imposed
on the \HMNS{} surface with a constant value up to $10$~ms and a time dependence
$t^{-1/2}$ \citep[e.g.,][]{pons:99} thereafter, with $t$ being the physical simulation
time. The electron neutrino and antineutrino luminosities have the same
magnitude, which we normalize to $2\times 10^{52}\,\text{erg s}^{-1}$ at 30~ms,
roughly matching the results of \citet{dessart:09}:
\begin{equation}
\label{eq:lnu_hmns} L_\nu = L_{\bar\nu} = 2\sqrt{3}\times
10^{52}\left[\max\left(1,\frac{t}{10\textrm{ ms}}\right)\right]^{-1/2}
\textrm{erg s}^{-1}.
\end{equation}
The neutrino and antineutrino temperatures
are different, however, chosen to be 4 and 5~MeV, respectively, which roughly
corresponds to the values found in proto-neutron stars in core-collapse
supernovae \citep[e.g.,][]{janka:01}. Thus, the mean neutrino energies from the
\HMNS{} are fixed at $\langle E_{\nu}\rangle = 12.6$~MeV and $\langle
E_{\bar{\nu}}\rangle = 15.8$~MeV (assuming a Fermi-Dirac distribution of
neutrinos, the mean energy is given by $\langle E_\nu \rangle =
3.151\,k_B\,T_\nu$, where $k_B$ is the Boltzmann constant and $T_\nu$ is the
neutrino temperature). These mean neutrino energies from the HMNS are broadly
consistent with results from numerical relativity simulations that include
neutrino transport \citep{foucart:16}. The \HMNS{} is transformed into a \BH{}
by switching the inner radial boundary from reflecting to absorbing and setting
the \HMNS{} neutrino luminosities to zero. The viscous stress is also set to
zero at this boundary when the \BH{} forms. See \citet{metzger:14} for more
details.

The initial condition for most models is an equilibrium torus obtained by
solving the Bernoulli equation with constant specific angular momentum and
electron fraction $Y_e = 0.1$ \citep[e.g.,][]{papaloizou:84,fernandez:13}. Outside the
torus, the computational domain is filled with a low-density ambient medium that
follows a power-law in radius, with an initial normalization $\sim 9$ orders of
magnitude below the maximum torus density ($\rho_\text{max} \sim 10^{11}\,\text{g cm}^{-3}$). The density floor inside $200$~km
decreases with time toward a constant asymptotic value of $10$~g~cm$^{-3}$ over
a timescale of $\sim 100$~ms. In addition, one model initializes the disk using
a snapshot from a general-relativistic NSBH simulation reported in
\citet{foucart:15}. Details about the mapping procedure can be found in
\citet{fernandez:17a}. Improvements to the code relative to \citet{metzger:14}
include the use of separate neutrino temperatures for disk self-irradiation, and
a correction to the weak interaction rates; details are provided in the
Appendix.

Passive tracer particles record thermodynamic quantities as a function of time,
and the resulting information is used as input for the nuclear reaction network
calculations (\S\ref{s:skynet_description}). The initial particle locations
are randomly sampled to follow the mass distribution in the disk. We
place 10,000 particles initially in all simulations. If there is a \BH{} at the
center, particles can fall into it, which then reduces the total number of
tracer particles. Fluid quantities are obtained at each time step from the grid
by linear interpolation. When a \HMNS{} is at the center, the reflecting
boundary for particles is placed one cell outside the inner radial boundary to
prevent particle trapping in a `trench' of small-magnitude negative radial
velocity that forms in the active cells adjacent to this boundary. The small
size of this innermost radial cell relative to the inner boundary radius
($\Delta r / r \simeq 1.8\%$) ensures that the effect on the particle dynamics
is minimal considering all other approximations being made.

Most hydrodynamic simulations are evolved up to about $10$~s of physical time
(\S\ref{s:investigated_models}), which is sufficient for r-process
nucleosynthesis that takes place within a few seconds. To obtain the final
abundances, however, the nuclear reaction network needs to be evolved for
tens of years, since some of the isotopes produced by the r-process have very
long half-lives. One caveat with this post-processing
approach is that the energy released by the nuclear reactions does not feed back
into the hydrodynamic evolution of the fluid
(with the exception of $\alpha$-recombination, which is accounted for in
the hydro evolution).
Nuclear heating may slightly
change the morphology of the ejecta at late times \citep[see, e.g., figure 8
of][]{fernandez:15} and influence specific features of the nucleosynthesis
indirectly via the amount of convection in the outflow \citep[][see also
\S\ref{s:final_abundances}]{wu:16}.
But the majority of heating due to the r-process and nuclear decays occurs
after particles have been ejected from the disk, so the dynamics is not
significantly altered by not including this heating in the hydrodynamic
simulation.

\subsection{Nuclear reaction network: \skynet}
\label{s:skynet_description}

We employ the nuclear reaction network \skynet\ for the r-process nucleosynthesis
calculations \citep{lippuner:17b}. For each thermodynamic trajectory
(\S\ref{s:flash_description}), we begin the reaction network evolution once the
temperature falls below 10~GK or reaches its maximum, if the maximum is less
than 10~GK. Since the composition is given by \NSE{} at temperatures above
10~GK, there is no need to start the reaction network evolution at higher
temperatures.

\skynet\ includes the specific viscous heating and neutrino heating/cooling
rates recorded by the thermodynamic trajectories, and adds to these a
self-consistent calculation of heating from nuclear reactions.
In addition to the neutrino heating/cooling rate, the associated rates of
neutrino interactions with free neutrons and protons are also given by the
thermodynamic trajectories and evolved in \skynet. The included reactions are electron (anti) neutrino
emission ($\text{p} + e^- \to \text{n} + \nu_e$, $\text{n} + e^+ \to \text{p} +
\bar{\nu}_e$) and neutrino absorption ($\text{n} + \nu_e \to \text{p} + e^-$,
$\text{p} + \bar\nu_e \to \text{n} + e^+$).

\skynet\ evolves the abundances of 7843 nuclides, ranging from free neutrons and
protons to \smce{^{337}Cn}, and includes over 140,000 nuclear reactions. The
strong reaction rates are taken from the JINA REACLIB database
\citep{cyburt:10}, but only the forward rates are used and the inverse rates are
computed from detailed balance. Spontaneous and neutron-induced fission rates
are taken from \citet{frankel:47}, \citet{panov:10}, \citet{mamdouh:01}, and
\citet{wahl:02}. Most of the weak rates come from \cite{fuller:82},
\cite{oda:94}, and \cite{langanke:00} whenever they are available, and otherwise
the REACLIB weak rates are used. The nuclear masses and partition functions used
in \skynet\ are taken from the WebNucleo XML file distributed with REACLIB,
which contains experimental data where available and finite-range droplet
macroscopic model \citep[FRDM, see, e.g.,][]{moller:16} data otherwise.

\subsection{Investigated models}
\label{s:investigated_models}

\newcommand{\0}{\phantom{0}}
\newcommand{\p}{\phantom{.}}

\begin{table}
\caption{List of investigated models. Columns from left to right show the model
name, the compact central object (CCO) type (\HMNS{} or \BH{}), mass $M_c$ of
the CCO, lifetime $\tau$ of the \HMNS{}, dimensionless spin $\chi$ of the
\BH{} (the \HMNS{s} all spin at $1.5$~ms), radius $R_d$ of the
initial disk density peak, and viscosity parameter $\alpha$.}
\label{tab:models}
\centering
\begin{tabular}{@{}l@{\ \ \ }cccccc@{}}
\hline
Model & CCO &  $M_{\rm c}$ & $\tau$ & $\chi$ & $R_d$ & $\alpha$  \\
      &     & ($M_\odot$)  & (ms)   &        & (km)  &           \\
\hline
H000  & HMNS & 3\p\0 & \0\00      & 0\p\0\0 & 50 & 0.03    \\
H010  &      &       &  \010      &         &    &         \\
H030  &      &       &  \030      &         &    &         \\
H100  &      &       &   100      &         &    &         \\
H300  &      &       &   300      &         &    &         \\
Hinf  &      &       & \0$\infty$ &         &    &         \\
\noalign{\smallskip}
B070  &   BH & 3\p\0 & \0\00      & 0.7\0   & 50 & 0.03    \\
B090  &      &       &            & 0.9\0   &    &         \\
\noalign{\smallskip}
BF15  &   BH & 8.1   & \0\00      & 0.86    & 55 & 0.03    \\
\noalign{\smallskip}
HinfNoVisc   & HMNS & 3\p\0 & \0$\infty$ & 0\p\0\0 & 50 & 0\p\0\0 \\
\hline \\
\end{tabular}
\end{table}

Table~\ref{tab:models} summarizes the properties of all investigated models. The
main focus of this study is the outflow from a disk around a \HMNS{} of variable
lifetime. Our baseline sequence follows the parameter choices of
\citet{metzger:14}: a \HMNS{}\ mass of $3\,M_\odot$ arising from an NSNS merger,
with an initial disk mass $M_d = 0.03\,M_\odot$ chosen as a representative case
of disk masses obtained in NSNS mergers \citep[e.g.,][]{hotokezaka:13}. We
prescribe the lifetime $\tau$ of the \HMNS{}\ to be 0, 10, 30, 100, or 300~ms,
after which the \HMNS{}\ collapses to a non-spinning \BH{}. We also run one
case, denoted by $\tau = \infty$, in which the \HMNS{} does not collapse. The
\HMNS{}\ models are denoted by H000, H010, H030, H100, H300, and Hinf, according
to their lifetime. Other disk parameters are: density peak radius $R_d = 50$~km,
viscosity parameter $\alpha=0.03$, constant initial entropy of $8\,k_B\ \text{baryon}^{-1}$,
constant initial $Y_e = 0.1$, and maximum evolution time of
$8.7$~s. These choices are motivated to be broadly compatible with results from
dynamical merger simulations \citep[e.g.,][]{ruffert:97,oechslin:06,foucart:16} and from studies of
angular momentum transport in fully-ionized accretion disks
\citep[e.g.,][]{davis:10}.

There are three important timescales in the problem:
the orbital time at the initial disk density peak,
\begin{equation}
\label{eq:t_orb}
t_{\rm orb}\simeq 3\left(\frac{R_d}{50\textrm{ km}} \right)^{3/2}
\left(\frac{3\,M_\odot}{M_c}\right)^{1/2}\textrm{ms},
\end{equation}
where $M_c$ is the mass of the central object (HMNS or BH); the initial thermal
time in the disk,
\begin{equation}
\label{eq:t_thermal}
t_{\rm th}\simeq 30\left(\frac{M_d}{0.03\,M_\odot}\right)\left(\frac{e_{i,d}}{10^{19}\textrm{ erg g}^{-1}}\right)
\left(\frac{2}{L_{\nu,52}} \right)\textrm{ms},
\end{equation}
where $e_{i,d}$ is the initial specific internal energy of the disk (a byproduct
of shock heating during the dynamical phase of the merger) and $L_{\nu,52}$ is a
typical neutrino luminosity from the disk in units of $10^{52}$~erg~s$^{-1}$
(neutrino cooling is in approximate balance with viscous heating at early
times); and the initial viscous time of the disk,
\begin{equation}
\label{eq:t_visc}
t_{\rm visc}\simeq 200\left(\frac{0.03}{\alpha}\right)\left(\frac{0.3}{H/R}\right)^2
\left(\frac{R_d}{50\textrm{ km}}\right)^{3/2}\left(\frac{3\,M_\odot}{M_c}\right)^{1/2}\textrm{ms},
\end{equation}

where $H/R$ is the height-to-radius ratio of the disk. Outflows driven primarily
by neutrino energy deposition are expected to be launched on the thermal
timescale (equation~\ref{eq:t_thermal}), whereas long-term outflows are launched
on the viscous timescale (equation~\ref{eq:t_visc}). The latter becomes longer
with time, as the disk spreads out and most of the mass in the disk resides at
an increasingly larger radius \citep[e.g.,][]{metzger:09c}. Therefore, the
length of the simulations has to be several viscous timescales, which translates
into thousands of orbits. We define convergence in mass ejection from the disk as
a saturation in the cumulative mass crossing some radius far away from the disk
($10^9$~cm). This way, we generally require $3000\,t_{\rm orb}$, or about $\sim 10$~s for
each simulation.

In order to examine the impact of the central compact object on the
nucleosynthesis, we evolve two additional models with spinning \BH{s} at the
center, following the approach of \citet{fernandez:15b}. The \BH{} mass is $M_c
= 3\,M_\odot$ in both cases and the dimensionless spin parameter $\chi = Jc/(GM_c^2)$
is  0.7 or 0.9, where $J$ is the \BH{} angular momentum, $c$ is the speed
of light, and $G$ is the gravitational constant. These models are labeled B070 and B090, according to their spin. Other
parameters are the same as in the \HMNS{}\ models. In addition, we investigate
the effect of the disk compactness (mass of the central compact object divided
by the disk density peak radius, $M_c/R_d$, which measures the strength of the gravitational
field) by evolving a model in which the initial condition is taken from a
snapshot of a general-relativistic simulation of a NSBH merger from
\citet{foucart:15}, including only the remnant accretion disk.
This model has a \BH{} mass of $8.1\,M_\sun$ and a disk
density peak radius of $55$~km, resulting in a gravitational potential that is
$2.5$ times stronger than our fiducial case (which better approximates the
remnant of an NSNS merger). The model is denoted by BF15, and the mapping
details are described in \citet{fernandez:17a}.

Finally, we evolve a test model that mirrors the case of a \HMNS{} with
$\tau=\infty$, but with the viscosity parameter set to zero, in order to
eliminate angular momentum transport and viscous heating. This model, denoted by
HinfNoVisc, experiences an outflow driven solely by neutrino heating (possibly
aided by nuclear recombination), and is evolved in order to compare results with
\citet{martin:15}. The model is evolved for a longer time ($14.5$~s) since mass
ejection converges more slowly with time relative to the viscous case.

\section{Results and discussion}
\label{s:results}

\subsection{Overview of disk evolution}
\label{sec:disk_evol}

\begin{figure*}
\includegraphics[width=\linewidth]{%
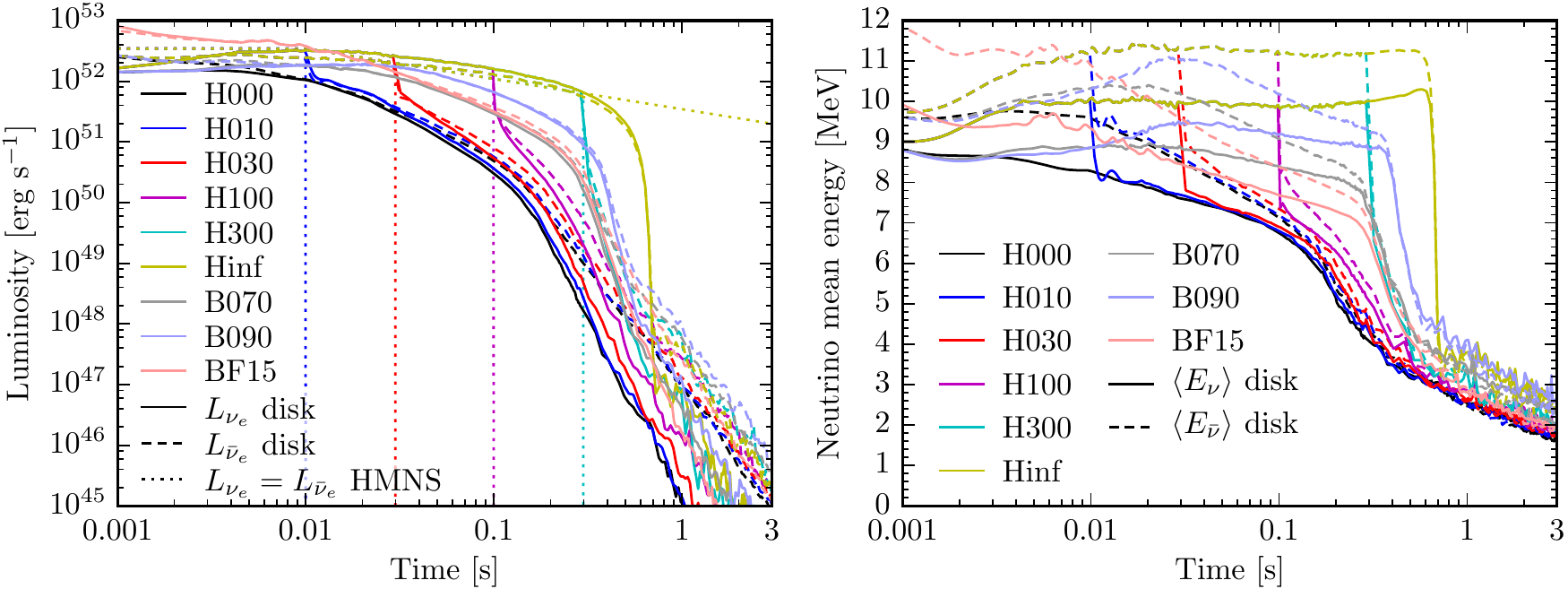}
\caption{\emph{Left:} Neutrino luminosities as a function of time for most
models studied in this paper (see Table~\ref{tab:models} for a summary). Shown
are electron neutrinos (solid lines) and electron antineutrinos (dashed lines)
emitted by the disk, as well as the imposed neutrino/antineutrino emission from
the \HMNS{} surface (dotted lines, equation \ref{eq:lnu_hmns}). Sharp drops in
the dotted lines mark the collapse of the \HMNS{} into a \BH{}. \emph{Right:}
Mean energies of electron neutrinos (solid lines) and electron antineutrinos
(dashed lines) emitted by the disk. The mean energies of neutrinos and
antineutrinos emitted by the \HMNS{} are fixed at 12.6~MeV and 15.8~MeV,
respectively.}
\label{fig:nu_lum}
\end{figure*}

The evolution of the disk and especially the neutrino interactions occurring in
the disk set the stage for the outflow and determine its properties. In this
section, we present a brief summary of the disk evolution, which was described
in detail in \citet{fernandez:13} and \citet{metzger:14}.

When a \HMNS{} is present, transport of angular momentum causes accreting
material to form a boundary layer around the reflecting stellar surface. The
outer regions of the disk expand on a thermal timescale
(equation~\ref{eq:t_thermal}) due to energy injection by neutrino heating and
viscous heating. Upon collapse of the \HMNS{} to a \BH{}, the boundary layer is
swallowed, and a rarefaction wave moves outward from the inner boundary,
quenching the thermal outflow \citep[cf.~Figure~3 of][]{metzger:14}. The disk
re-adjusts on a thermal timescale, and joins the evolutionary path of a \BH{}
accretion disk, which changes on a viscous timescale (equation~\ref{eq:t_visc}).

The high densities achieved on the equatorial plane of \HMNS{} disks ($\sim 10^{11}\,\text{g cm}^{-3}$) are enough
to locally trap neutrinos, resulting in two emission hot spots at
mid latitudes and adjacent to the \HMNS{} surface. The high densities in the
midplane also result in shadowing of the outer disk, with neutrino irradiation
concentrated at latitudes $\sim 30^\circ$ away from the equator
\citep[cf.~Figure~2c of][]{metzger:14}. This general neutrino irradiation geometry is also found
when better (Monte Carlo) radiation transport is performed on snapshots of
\HMNS{} disk models \citep{richers:15}.

The high densities near the boundary
layer cause the electron fraction in the inner regions of the \HMNS{} disk ($r <
10^7$~cm, and within $\sim 30^\circ$ from the equatorial plane) to remain low ($Y_e
\sim 0.1-0.2$) relative to the outer regions. Away from the midplane, the weak
interaction timescale becomes shorter, and material that enters the boundary
layer at high latitude reaches $Y_e \sim 0.5$ within an orbital timescale, as it
transits through the hot spot in neutrino emission.
Material ejected within
$\sim 45^\circ$ of the rotation axis has therefore high electron fraction
relative to the rest of the outflow. At $t = t_{\rm th}$, most of the disk
material within $r = 10^7$~cm is close to beta equilibrium, with $Y_e \sim 0.2-0.3$
in the outer regions. As the disk
approaches a viscous timescale, the density in the inner disk gradually decreases,
resulting in two effects: (1) the equilibrium $Y_e$ of the inner disk increases as the
degeneracy of the material decreases, and (2) the strength of the weak
interactions decreases, until weak interactions become slow relative to the
viscous time \citep[e.g.,][]{metzger:09c}. These two effects combine to leave
material close to the \HMNS{} with $Y_e\sim 0.4$. As long as the \HMNS{} does
not collapse, most of this material will eventually be ejected.

The \BH{} accretion disk is such that the inner regions ($r < 10^7$~cm) also
approach beta equilibrium, but most of the material that reaches high $Y_e$ is
accreted onto the \BH{} \citep[cf.~Figure~6 of][]{fernandez:13}. The fraction of
the high-$Y_e$ material that is ejected is a function of the spin of the \BH{}, since a
smaller \ISCO{} slows down accretion and allows material to reach regions where
weak interactions are stronger, before being ejected \citep{fernandez:15b}. Upon
collapse of the \HMNS{} into a \BH{}, a low density funnel of width $\sim 45^\circ$
around the rotation axis is created as material that has not yet been unbound is
swallowed. Boundary layer material reaches the highest $Y_e$ as it crosses the
emission hot spots of the \HMNS{} disk before being ejected, hence formation of
the \BH{} precludes further ejection of material with $Y_e \sim 0.5$. For the
disk masses simulated, the material becomes optically thin within a few orbits
if a \BH{} is at the center, therefore hot spots also disappear upon collapse
of the \HMNS{}.

\subsection{Neutrino emission}
\label{sec:nu_irr}

The total electron neutrino and antineutrino luminosities from the disk and the
\HMNS{} (if present) as a function of time are shown in the left panel of
Figure~\ref{fig:nu_lum}. The disk luminosities are very similar to the
luminosity from the \HMNS{} as long as the latter does not collapse, because the
disk absorbs and re-radiates the neutrinos emitted by the \HMNS{}. When the
\HMNS{} collapses, the disk luminosities undergo a sudden decrease, and then
follow a broken powerlaw as the disk re-adjusts and continues to evolves on the
viscous timescale. Thereafter, $L_{\bar{\nu}_e}$ starts to dominate over
$L_{\nu_e}$ in all models because the initial neutron-rich ($Y_e = 0.1$)
composition of the disk causes weak interactions to leptonize it, with more
positron captures than electron captures (the gradual drop in density over the
viscous timescale increases the equilibrium $Y_e$ of the disk;
\S\ref{sec:disk_evol}). As long as the \HMNS{} is present, the disk neutrino
luminosities are approximately the same in all cases, consistent with
re-radiation of the \HMNS{} luminosities, and the different models are separated
by a shift in time depending on the lifetime of the \HMNS{}.

The mean neutrino energies emitted by the disk are shown in the right panel of
Figure~\ref{fig:nu_lum}. Given the low initial abundance of protons, the optical
depth for antineutrinos is initially low, and antineutrino energies are higher
by $\sim 10\%$ than neutrino energies until $t\sim t_{\rm visc}$, by which time
the derease in disk density has gradually lifted the degeneracy and weak
interactions have driven the inner disk close to $Y_e \sim 0.4$. The
neutrino/antineutrino energies are the same between \HMNS{} models until the
\HMNS{} collapses, each following the same general evolution pattern but shifted
in time. Since the disk mostly re-radiates neutrinos from the \HMNS{}, the mean
neutrino energies drop sharply when the \HMNS{} collapses.

In the models that start out with a \BH{} at the center (B070, B090, BF15, and
H000, which has $\chi = 0$), a more rapidly spinning \BH{} results in higher
neutrino luminosities and mean neutrino energies. This occurs because larger
spins are associated with smaller \ISCO{} radii. Hence the disk material can
convert more gravitational energy into thermal energy, resulting in more
intense neutrino emission with higher mean energies. In model BF15, the central
\BH{} is more massive ($8.1\,M_\odot$ compared to $3\,M_\odot$ in H000, B070,
and B090), which results in a larger \ISCO{} radius. The disk has nearly the
same density peak radius as the other models, but the initial condition is not
in equilibrium. Therefore, accretion proceeds more intensely at early times than
in the other \BH{} models, speeding up the disk evolution despite its slightly
longer initial viscous time (smaller $H/R$ in equation~\ref{eq:t_visc}, which
overcomes the effect of a larger \BH{} mass).

\subsection{Ejecta properties}
\label{sec:ejecta_properties}

\begin{table*}
\caption{Summary of nucleosynthesis results. $N_\text{ej}$ is the number of
tracer particles that reach a radius $r=10^9$~cm by the end of the hydrodynamic
simulation (every simulation starts with 10,000 particles); $M_\text{ej}$ is the
total ejected mass in $10^{-3}\,M_\odot$ at the same radius;
$M(Y_e\!\leq\!0.25)$ is the ejected mass with $Y_{e,\text{5GK}} \leq 0.25$;
$M_{\nu\text{-driv}}$ is the amount of ejected mass that is driven by neutrino
interactions; $[Y_\text{1st}/Y_\text{2nd}]$ is the log$_{10}$ of the ratio of the first
r-process peak to the second r-process peak, normalized to the solar value, see
\cref{eq:peak_ratio} for details; $[Y_\text{RE}/Y_\text{2nd}]$ and
$[Y_\text{3rd}/Y_\text{2nd}]$ are the same quantities for the rare-earth and
third peaks; $\langle X_\text{La}\rangle$ and $\langle X_\text{Ac}\rangle$ are
the lanthanide and actinide mass fractions averaged over all ejecta particles;
and $\epsilon_\text{tot}$~1~d and $\epsilon_\text{tot}$~7~d are the total
heating rates of the entire ejecta at 1 and 7 days, respectively.}
\label{tab:ejecta}
\setlength{\tabcolsep}{3.9pt}
\begin{tabular}{@{}lccccccccccc@{}}
\hline
Model & $N_\text{ej}$ & $M_\text{ej}$ & $M(Y_e\!\leq\!0.25)$ & $M_{\nu\text{-driv}}$ &
$[Y_\text{1st}/Y_\text{2nd}]$ & $[Y_\text{RE}/Y_\text{2nd}]$ &
$[Y_\text{3rd}/Y_\text{2nd}]$ & $\langle X_\text{La} \rangle$ &
$\langle X_\text{Ac} \rangle$ & $\epsilon_\text{tot}$ 1 d &
$\epsilon_\text{tot}$ 7 d \\
& & ($M_{-3\odot}$) & ($M_{-3\odot}$) & ($M_{-3\odot}$) & & & & & & (erg~s$^{-1}$) &
(erg~s$^{-1}$) \\
\hline
\noalign{\smallskip}
H000 & \0527 & \01.8   & 1.4\0\0\0 & \00.013    & $-1.3$\0\0\0 & $-0.28$  & $-0.30$  & $4.6 \!\times\! 10^{-2}$ & $6.4 \!\times\! 10^{-3}$ & $9.0 \!\times\! 10^{40}$ & $1.4 \!\times\! 10^{40}$ \\
H010 & \0557 & \01.9   & 1.1\0\0\0 & \00.027    & $-1.1$\0\0\0 & $-0.40$  & $-0.50$  & $3.3 \!\times\! 10^{-2}$ & $2.0 \!\times\! 10^{-3}$ & $8.9 \!\times\! 10^{40}$ & $1.3 \!\times\! 10^{40}$ \\
H030 & \0989 & \03.3   & 0.83\0\0  & \00.20\0   & $-0.64$\0\0  & $-1.0$\0 & $-1.2$\0 & $5.1 \!\times\! 10^{-3}$ & $2.9 \!\times\! 10^{-4}$ & $1.2 \!\times\! 10^{41}$ & $1.5 \!\times\! 10^{40}$ \\
H100 &  2408 & \07.8   & 0.52\0\0  & \01.3\0\0  & $-0.0053$    & $-0.91$  & $-1.2$\0 & $2.1 \!\times\! 10^{-3}$ & $1.7 \!\times\! 10^{-4}$ & $1.8 \!\times\! 10^{41}$ & $1.4 \!\times\! 10^{40}$ \\
H300 &  5610 &  18\p\0 & 0.67\0\0  & \06.4\0\0  & $+0.25$\0\0  & $-0.88$  & $-1.2$\0 & $1.1 \!\times\! 10^{-3}$ & $6.7 \!\times\! 10^{-5}$ & $3.3 \!\times\! 10^{41}$ & $3.2 \!\times\! 10^{40}$ \\
Hinf &  9587 &  30\p\0 & 0.69\0\0  & $28^\ast$\p\0\0 & $+0.41$\0\0  & $-0.86$  & $-1.1$\0 & $7.1 \!\times\! 10^{-4}$ & $4.2 \!\times\! 10^{-5}$ & $5.2 \!\times\! 10^{41}$ & $5.4 \!\times\! 10^{40}$ \\
\noalign{\smallskip}
B070 &  1465 & \05.4   & 1.8\0\0\0 & \00.022    & $-0.73$\0\0  & $-0.65$  & $-0.67$  & $1.3 \!\times\! 10^{-2}$ & $1.6 \!\times\! 10^{-3}$ & $2.0 \!\times\! 10^{41}$ & $2.4 \!\times\! 10^{40}$ \\
B090 &  2363 & \07.9   & 1.6\0\0\0 & \00.070    & $-0.54$\0\0  & $-0.77$  & $-0.80$  & $7.6 \!\times\! 10^{-3}$ & $9.7 \!\times\! 10^{-4}$ & $2.7 \!\times\! 10^{41}$ & $2.6 \!\times\! 10^{40}$ \\
\noalign{\smallskip}
BF15 & \0910 & \04.9   & 0.011\0   & \00.022    & $-0.26$\0\0  & $-1.4$\0 & $-1.3$\0 & $1.4 \!\times\! 10^{-3}$ & $7.7 \!\times\! 10^{-5}$ & $7.8 \!\times\! 10^{40}$ & $6.7 \!\times\! 10^{39}$ \\
\hline \\[-4pt]
\multicolumn{12}{@{}p{\textwidth}@{}}{$^\ast$ Since the \HMNS{} persists forever
in model Hinf, virtually all trajectories experience significant neutrino
interactions and our method becomes inadequate to isolate the component of the
ejecta that is driven by neutrino interactions.} \\
\end{tabular}
\end{table*}

\begin{figure}
\includegraphics[width=\columnwidth]{%
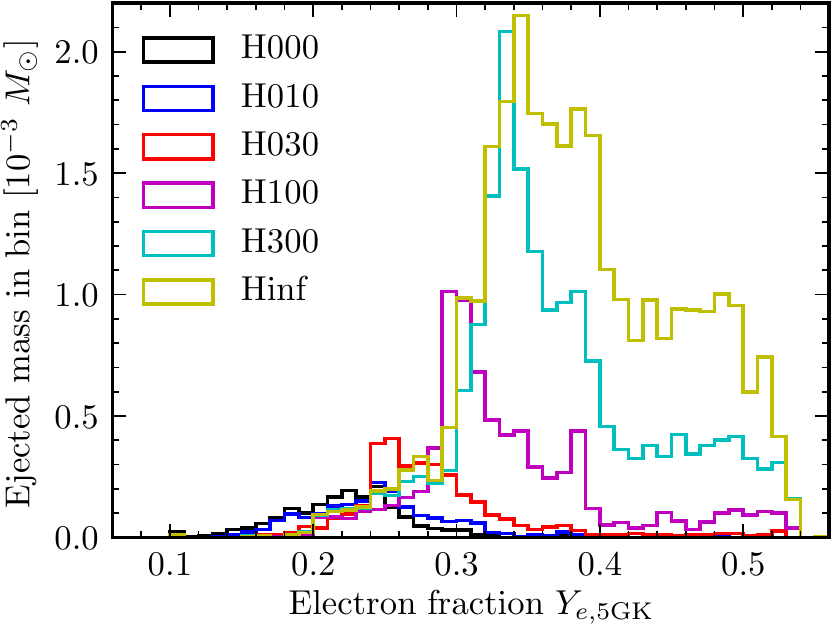}
\caption{Mass distribution of the ejecta electron fraction for \HMNS{} models at
the time when the temperature is $\geq 5$~GK for the last time for each tracer
particle.}
\label{fig:ye_hist_hmns}
\end{figure}

In order to associate the thermodynamic properties of the disk with the
nucleosynthesis outcome from each trajectory, we use the value of the electron
fraction ($Y_{e,\text{5GK}}$) and specific entropy ($s_\text{5GK}$) at the last time when
the temperature of the tracer particle drops below $5$~GK. For the rare cases in
which the temperature of the trajectory is always below $5$~GK, we use the
initial values of $Y_e$ and $s$ for $Y_{e,\text{5GK}}$ and $s_\text{5GK}$,
respectively. Once the temperature drops below approximately $5$~GK, the
composition moves out of \NSE{} and a full network evolution is required to
evolve the abundances. Therefore $Y_{e,\text{5GK}}$ and $s_\text{5GK}$ are the
initial conditions for nucleosynthesis. Note that the reaction network evolution
starts when the temperature drops below 10~GK, but \skynet\ can also evolve the
composition while \NSE{} holds.

The distribution of mass ejected as a function of $Y_{e,\text{5GK}}$ is shown in
Figure~\ref{fig:ye_hist_hmns} for all \HMNS{} models with non-zero viscosity. A
trajectory is considered to have been ejected when it crosses the surface $r =
10^9$~cm. There is a strong correlation between the \HMNS{} lifetime and both
the amount of mass ejected and the mean $Y_{e,\text{5GK}}$ of the distribution
\citep{metzger:14}. The disk ejecta ranges from 6 to nearly 100\% of the
initial disk mass. The longer the \HMNS{} lives, the longer the disk material is
subject to strong neutrino heating, which combines with viscous heating and
nuclear recombination to eject material on the viscous timescale. A longer \HMNS{}
lifetime also allows more material from the inner disk to be ejected instead of
being swallowed by the \BH{}. That material from the inner disk reaches beta
equilibrium and hence its ejection results in a higher
mean electron fraction. \Cref{tab:ejecta} shows the number of ejected particles
(out of the initial 10,000 particles in each model) and the total ejected mass.
Also shown is the amount of mass ejected with $Y_{e,\text{5GK}} \leq 0.25$, which
is neutron-rich enough to robustly make the full r-process \citep[see next
section and, e.g.,][]{lippuner:15}.

We note that the amount of mass ejected with $Y_{e,\text{5GK}} \leq 0.25$ is
roughly constant between $(0.5-0.8)\times 10^{-3}\,M_\odot$ once the \HMNS{}
lives for 30~ms or longer, despite the total ejecta mass differing by an order
of magnitude. This is the result of two competing effects: a longer \HMNS{}
lifetime increases the total ejecta mass, but it also increases the average
electron fraction of the ejecta, thus reducing the fraction of the ejected mass
that has $Y_{e,\text{5GK}} \leq 0.25$. These two effects counteract each other,
leaving the absolute amount of mass ejected with $Y_{e,\text{5GK}} \leq 0.25$
roughly constant.

\begin{figure}
\includegraphics[width=\columnwidth]{%
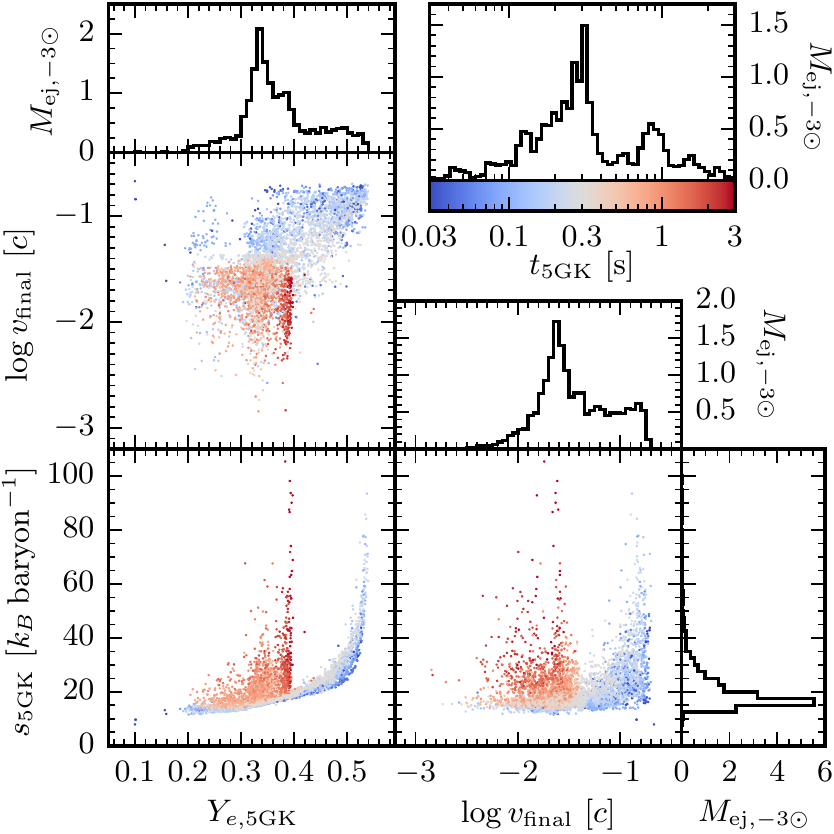}
\caption{Properties of the ejecta of model H300 (central \HMNS{} that collapses after 300 ms). $Y_{e,\text{5GK}}$ and
$s_\text{5GK}$ are the electron fraction and specific entropy at the time $t_\text{5GK}$,
when the temperature of the trajectory is 5~GK for the last time, while
$v_\text{final}$ is the final velocity of the trajectory. The scatter plot
points are color coded by $t_\text{5GK}$, which marks the nucleosynthesis
starting time. Blue and gray points ($t_\text{5GK} \lesssim 300$~ms) start their
nucleosynthesis while the \HMNS{} is present, and are thus influenced by the
neutrino irradiation from the \HMNS{}. The strong correlation between electron
fraction and entropy is similar to what is obtained in a neutrino-driven
outflow. The red points ($t_\text{5GK} \gtrsim 300$~ms) start nucleosynthesis
after the \HMNS{} has collapsed to a \BH{}. Hence this component is subject
primarily to the action of viscous processes and nuclear recombination in the
disk. See the text for details. $M_{\text{ej},-3\odot}$ means the amount of
ejecta mass in units of $10^{-3}\,M_\odot$.}
\label{fig:scatter}
\end{figure}

The thermodynamic properties of the ejecta for model H300 (\HMNS{} with lifetime
$\tau = 300$~ms) are illustrated in \cref{fig:scatter} through the electron
fraction $Y_{e,\text{5GK}}$, specific entropy $s_{\text{5GK}}$, final velocity
$v_\text{final}$, ejecta mass, and the time $t_{\text{5GK}}$ when the
temperature is $5$~GK for the last time. Two ejecta components stand out from
the scatter plot of $Y_{e,\text{5GK}}$ versus $s_{\text{5GK}}$. The larger
component is ejected before the \HMNS{} collapses, i.e.\ $t_\text{5GK} \lesssim
300$~ms, and it exhibits a tight correlation between the entropy and electron
fraction up to $Y_{e,\text{5GK}} \sim 0.5$. This is indicative of a
neutrino-driven wind, and indeed we would expect the asymptotic $Y_e$ to be
$\sim 0.55$ based on the neutrino properties shown in \cref{fig:nu_lum}
\citep{qian:96}. Note that the vast majority of the ejecta with
$Y_{e,\text{5GK}} \gtrsim 0.4$ or $v_\text{final} \gtrsim 0.03\,c$ is part of
this early wind-like ejecta, with a much smaller group of particles extending to
low velocities and low electron fractions.

The second component is ejected after the \HMNS{} has collapsed to a \BH{},
i.e.\ $t_\text{5GK} \gtrsim 300$~ms, and it has only a weak correlation between
$Y_{e,\text{5GK}}$ and $s_\text{5GK}$. This component is associated with mass
ejection as the disk reaches the advective state (very weak or no neutrino
cooling/heating) and is driven primarily by heat injection from angular momentum transport processes
and nuclear recombination \citep{metzger:14}. Mass ejection in this state is
accompanied by vigorous convective activity in the disk.

In order to quantitatively disentangle the wind-like ejecta component seen in
\cref{fig:scatter} from the other component, we compute the contributions to
$s_\text{5GK}$ arising from neutrino and viscous
heating. While all trajectories experience some degree of viscous heating, only
a subset of tracer particles experience an entropy change due to
neutrino heating that is larger than $0.1\ k_B\ \text{baryon}^{-1}$ (ignoring any neutrino cooling). And that subset exactly exhibits the tight, boomerang-shaped
correlation between $\ye$ and $s_\text{5GK}$ that the early, wind-like ejecta
exhibits (cf.~the blue and gray dots in the lower left panel of \cref{fig:scatter}).
The mass of this neutrino-driven ejecta component is shown in
\cref{tab:ejecta} under $M_{\nu\text{-driv}}$. This component is absent for
\HMNS{} lifetimes shorter than 30~ms and also in the \BH{} models, but once the
\HMNS{} lifetime becomes longer, the neutrino-driven component becomes as
large or even the dominant fraction of the total ejecta (such as in model Hinf, in
which almost the entire ejecta is neutrino-driven). For very long-lived
\HMNS{}s all trajectories experience some degree of neutrino interactions, and
it becomes difficult to distinguish the wind-like from the viscous component. We
emphasize that even the neutrino-driven component experiences significant
viscous heating and the entropy change due to viscous heating can be of the same order as
that due to neutrino absorption.

One interesting feature of the late-time component is the sharp upper limit of
$Y_{e,\text{5GK}} \sim 0.38 - 0.40$, regardless of entropy. These trajectories
experience late-time fall back due to large convective eddies in the disk. They
get sucked deep inside the disk where the density is much higher than in the
outflow, and then they are ejected again almost immediately. This creates a
spike in their density profile that results in significant heating, as evidenced
by the fact that they all have $T \geq \text{5 GK}$ at $t \sim 2$~s. However,
before this late-time heating occurs, r-process nucleosynthesis has already
taken place in these trajectories, and all free neutrons have been captured onto
seed nuclei. Thus, the composition before the heating spike consists of heavy
elements with $\beta$-decay half-lives of milliseconds to seconds. These
elements decay and raise the overall electron fraction of the material to $Y_e
\sim 0.38 - 0.40$, which is the characteristic $Y_e$ at $1 - 3$ seconds after
neutron exhaustion for the r-process, for a wide range of initial $Y_e$. The
late-time heating then simply pushes the material back into \NSE{}, but the
electron fraction remains unchanged. The resulting entropy depends on the amount
of heating received by each trajectory, as determined by how far the material
falls back into the disk. This class of trajectories therefore ends up with
electron fractions $Y_{e,\text{5GK}} \sim 0.38 - 0.40$ and nucleosynthesis start
times of $t_\text{5GK} \sim 2$~s, with uncorrelated entropies.

\subsection{Nucleosynthesis}

\subsubsection{Final abundances}
\label{s:final_abundances}

\begin{figure}
\includegraphics[width=\columnwidth]{%
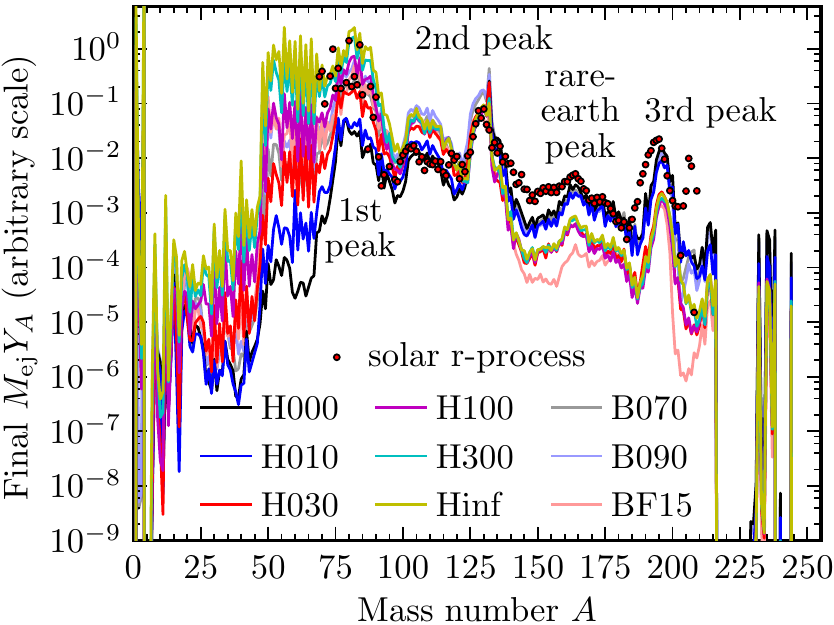}
\caption{Final trajectory-averaged abundances as a function of mass number,
scaled by the total ejecta mass, for all models with non-zero viscosity. The
observed solar r-process abundances \citep{arnould:07} are scaled to match the
second peak of the \HMNS{} models at $A = 130$ (none of the abundances from our
models have been scaled).}
\label{fig:abund}
\end{figure}

The mass-averaged composition of the ejecta for all models with non-zero
viscosity is shown in \cref{fig:abund}. The abundances are multiplied by the
total ejecta mass to emphasize their relative contributions to the different
r-process regions. Models H000 and H010 (prompt non-spinning \BH{} and
shortest-lived \HMNS{}, respectively) agree most closely with the Solar System
r-process abundances \citep{arnould:07}, which have been scaled to match the
second peak at $A = 130$ (the abundances from our models have not been scaled).
The abundances around the third r-process peak in these two models approach the
solar values, whereas in all other models production of the third peak is too
low compared to solar. H000 and H010 also have the best agreement with the solar
rare-earth peak around $A \sim 165$. While these two models under-produce the
first r-process peak ($A \sim 80$), they agree rather well with the feature
around $A \sim 100$, in contrast to all other models which over-produce it.

While the good agreement between models H000/H010 and the solar r-process
abundances could be taken as an indication of short \HMNS{s} lifetimes being
more common, one has to keep in mind that \cref{fig:abund} assumes that the
entire second solar r-process peak is due to the disk outflow. Other sources
such as the dynamical ejecta from NSNS/NSBH mergers and core-collapse supernovae
can also produce significant amounts of r-process elements. The expected abundance
patterns are weighted toward the third peak for the dynamical ejecta
\citep[e.g.,][]{goriely:11,wanajo:14,roberts:16b} and toward the first peak for
core-collapse supernovae \citep[e.g.,][]{wanajo:13,shibagaki:16,vlasov:17}. The solar
r-process abundance is thus the outcome of the contribution from each source
weighted by their rate and yield per event.

In all models, the third peak is shifted to slightly higher mass numbers, which
is a well-known shortcoming of the FRDM mass model
\citep[e.g.,][]{mendoza:15,mumpower:16}. We also see an abundance spike at $A =
132$ in all models. This spike is due to some trajectories experiencing
late-time heating that photodissociates neutrons from synthesized heavy
elements. This results in additional neutron capture and a pile up of material at
the doubly magic nucleus \smce{^{132}Sn} ($N = 82$ and $Z = 50$). \citet{wu:16}
also observed this phenomenon and described it in detail.

The models with longer \HMNS{} lifetimes have less neutron-rich ejecta
(Figure~\ref{fig:ye_hist_hmns}) and hence synthesize a greater fraction of first
peak material. Once the \HMNS{} lifetime is longer than 100~ms, the first
peak ($70 \leq A \leq 90$) is over-produced with respect to the solar values,
when the abundances are normalized to the second peak. Again, we emphasize that
the r-process yield from disk outflows is complementary to that from the
dynamical ejecta, which tends to produce more neutron-rich nuclei.

We quantify the relative contribution of each model to the different regions of
the r-process distribution by computing average abundances around the peaks and
normalizing them to the solar values. The abundance of the second peak
$Y_\text{2nd}$ is computed as the sum of the abundances in the range $125 \leq A
\leq 135$, excluding $A = 132$ to avoid the spike at that mass number. For the
first peak abundance $Y_\text{1st}$, we use the sum of abundances in the range
$70 \leq A \leq 90$. For the rare-earth peak $Y_\text{RE}$, we use $160 \leq A
\leq 166$ and for the third peak we use $186 \leq A \leq 203$. The quantity
$[Y_\text{1st}/Y_\text{2nd}]$ shown in \cref{tab:ejecta} is defined as
\begin{align}
[Y_\text{1st}/Y_\text{2nd}] = \log_{10}\frac{Y_{\text{1st}}}{Y_{\text{2nd}}} -
\log_{10}\frac{Y_{\text{1st},\odot}}{Y_{\text{2nd},\odot}}, \label{eq:peak_ratio}
\end{align}
where $Y_{\text{1st},\odot}$ and $Y_{\text{2nd},\odot}$ are the abundances of
the third and second peak as observed in the solar system, respectively. The
same procedure is used to compute $[Y_\text{RE}/Y_\text{2nd}]$ and
$[Y_\text{1st}/Y_\text{2nd}]$. Using the solar r-process abundances from
\citet{arnould:07}, we find $\log Y_{\text{1st},\odot}/Y_{\text{2nd},\odot} =
+1.3$, $\log Y_{\text{RE},\odot}/Y_{\text{2nd},\odot} = -1.1$, and $\log
Y_{\text{3rd},\odot}/Y_{\text{2nd},\odot} = -0.42$, which we use to normalize
the values shown in \cref{tab:ejecta}.

The different peak ratios shown in \cref{tab:ejecta} quantify the trends
apparent in \cref{fig:abund}. For models H000 and H010, the rare-earth and third
peaks are under-produced by only a third to one half of an order of magnitude.
But the first peak is under-produced by slightly more than an order of
magnitude compared to the second peak in those models. As we go to longer \HMNS{} lifetimes, the
rare-earth and third peaks are under-produced by about an order of magnitude
regardless of the \HMNS{} lifetime. At the same time, the first peak increases from an
under-production of $2/3$ of an order of magnitude at $\tau = 30$~ms to an
over-production of a factor of 2.6 at $\tau = \infty$.

The spinning \BH{} models under-produce the first, rare-earth, and third peaks
in roughly the same amounts, namely between about half to $3/4$ of an order of
magnitude compared to the second peak. The compact disk model BF15
under-produces the third peak by a similar amount as the \HMNS{} models with
long lifetimes. But it also has the lowest rare-earth peak abundance relative to
solar, with an under-production factor of about 1.5 orders of magnitude. The
first peak, on the other hand, is only under-produced by a factor of $\sim 2$.

\subsubsection{BH spin mimicking HMNS lifetime}

\begin{figure}
\includegraphics[width=\columnwidth]{%
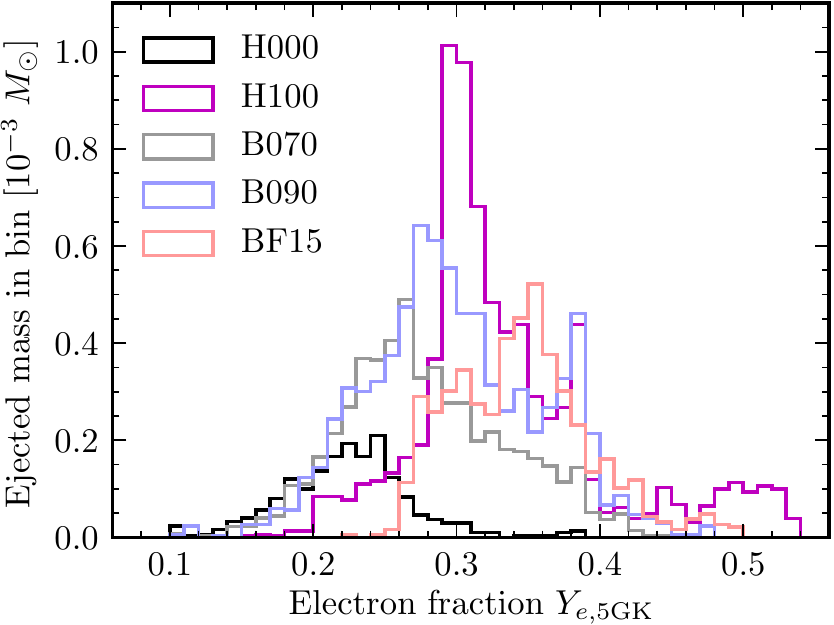}
\caption{Mass histograms of electron fraction for the ejecta from \BH{} models.
We include the \HMNS{} model H100 for comparison.}
\label{fig:ye_hist_bh}
\end{figure}

\citet{metzger:14} proposed using the relative amount of blue optical emission
in the kilonova as an observational test of the \HMNS{} lifetime, given that the
latter correlates with the amount of high-$Y_e$ material ejected in the disk
outflow. Given that high-$Y_e$ material also correlates with \BH{} spin
\citep{fernandez:15b}, the kilonova signature of the two types of central
objects can overlap. Here we investigate whether the nucleosynthesis signature
offers additional information that can break the degeneracy. We consider the
\BH{} models H000, B070, and B090, which have the same mass and varying spins
$\chi = 0$, 0.7, and 0.9, respectively, as well as model BF15, which has a
larger mass and $\chi = 0.86$. \Cref{fig:ye_hist_bh} shows the electron fraction
distributions of these \BH{} models.

While a larger \BH{} spin has a similar overall effect on the disk ejecta
composition as a longer \HMNS{} lifetime, even a spin $\chi = 0.9$ does not
reach the same amount of ejecta mass and average value of $Y_{e,\text{5GK}}$ as
a \HMNS{} with a lifetime $\tau = 100$~ms (cf.\ \cref{fig:ye_hist_bh}). At best,
a rapidly spinning \BH{} can mimic a \HMNS{} of modest lifetime. This can also
be seen in \cref{fig:abund} and in the peak ratio values in \cref{tab:ejecta}.
Models B070 and B090 have values of $[Y_\text{1st}/Y_\text{2nd}]$ that are
similar or slightly smaller than in model H030, while the values of
$[Y_\text{RE}/Y_\text{2nd}]$ and $[Y_\text{3rd}/Y_\text{2nd}]$ from models B070
and B090 fall between those from models H010 and H030. Thus, a \BH{} with a spin
$\chi = 0.7-0.9$ produces a disk outflow with a similar final abundance pattern
as a \HMNS{} with a lifetime of $\tau \sim 15 - 20$~ms. Therefore, the fact that
we assume a non-spinning \BH{} after collapse of the \HMNS{} only has a very
small impact on the nucleosynthesis. Model BF15, on the other hand, looks more
like a \HMNS{} with a lifetime $\tau \sim 60$~ms, judging from the
under-production of the first and third peaks relative to the second peak.
Nonetheless, the rare-earth peak is under-produced by almost half an order of
magnitude more in the disk ejecta by BF15 than by the \HMNS{} models with long
lifetimes.

\subsubsection{Lanthanides, actinides, and heating rates}

\begin{figure*}
\includegraphics[width=\linewidth]{%
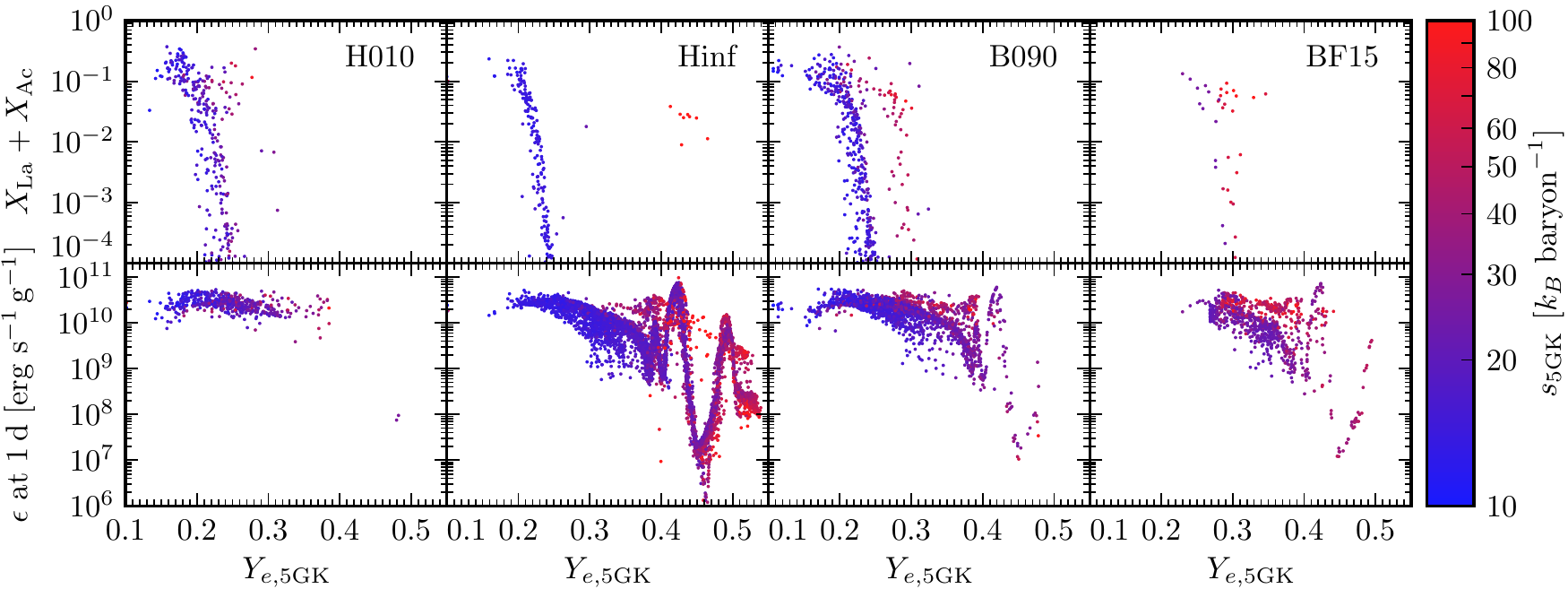}
\caption{\emph{Top:} Scatter plots of final lanthanide and actinide mass
fraction in each trajectory as a function of the electron fraction
$Y_{e,\text{5GK}}$, for selected models. Points are color-coded by the initial
entropy $s_\text{5GK}$. \emph{Bottom:} Radioactive heating rate $\epsilon$ at
one day as a function of $Y_{e,\text{5GK}}$, also color-coded by
$s_\text{5GK}$.}
\label{fig:laac_heat}
\end{figure*}

The energy released by the radioactive decay of heavy elements synthesized by
the r-process can power an optical or infrared transient called a kilonova
(also called macronova in the literature; \citealt{li:98,kulkarni:05,
metzger:10b,roberts:11,kasen:13, tanaka:13, grossman:14, metzger:17,
wollaeger:17}).
The two most important microphysical
components that determine the light curve and spectrum of kilonovae are the
opacity of the material and the radioactive heating rate. Lanthanides
($58 \leq Z \leq 71$) and actinides ($90 \leq Z \leq 103$) have open f-shells,
which gives them a very complex atomic line structure that leads to broadband
opacities that are more than an order of magnitude larger than opacities from
iron-group elements \citep{kasen:13,tanaka:13,fontes:15}.

The trajectory-averaged lanthanide $\langle X_\text{La} \rangle$ and actinide
$\langle X_\text{Ac}\rangle$ mass fractions of the ejecta for all models are
summarized in \cref{tab:ejecta}. For \HMNS{} lifetimes $\tau \leq 10$~ms, the
lanthanide mass fraction is a few times $10^{-2}$, which will result in
opacities about an order of magnitude larger than iron group opacities
\citep[see Figure 10 in][]{kasen:13}. Note that the actinide mass fractions are
on the order of $10^{-3}$, which is still a significant contribution to the
opacity. Once the \HMNS{} lifetime is longer than about $10$~ms, the lanthanide
fraction $\langle X_\text{La}\rangle \sim 10^{-3}$ monotonically decreases as
the lifetime increases. In these models, the lanthanides and actinides will
increase the opacity only by a factor of a few relative to iron group opacities.
In the \BH{} models, increasing the spin from 0.7 to 0.9 reduces the
lanthanide and actinide mass fractions by roughly a factor of two from around
$10^{-2}$. The compact disk model BF15 has very similar lanthanide and actinide
mass fractions as H300. These results apply to the disk outflow component alone;
the color of a kilonova depends also on the spatial distribution and composition
of the dynamical ejecta, which tends to be more neutron-rich and hence may
contain significant amounts of lanthanides and actinides.

\Cref{fig:laac_heat} shows the combined lanthanide and actinide mass fraction of
each ejected trajectory as a function of $Y_{e,\text{5GK}}$ for models H010,
Hinf, B090, and BF15. For most trajectories, the lanthanide and actinide
fraction plummets as the initial electron fraction increases from $0.2$ to
$0.25$ \citep[see also][]{lippuner:15,kasen:15}. Particles with $Y_{e,\text{5GK}} \leq
0.25$ have low entropies, because a higher entropy requires either significant
neutrino heating, which increases $Y_e$, or viscous heating over timescales
comparable to the thermal time, which gives weak interactions enough time to
increase $Y_e$ toward its equilibrium value. This correlation between
$Y_{e,\text{5GK}}$ and $s_\text{5GK}$ at $Y_{e,\text{5GK}} \lesssim 0.3$ can be
seen in \cref{fig:scatter}.

\begin{figure}
\includegraphics[width=\columnwidth]{%
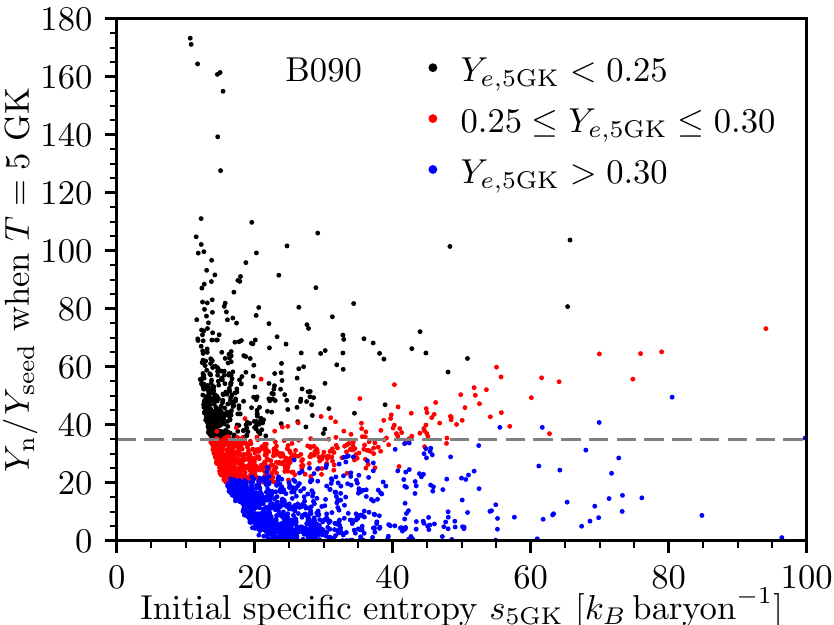}
\caption{Neutron-to-seed ratio at the time when the temperature $T \geq 5$~GK
for the last time, versus the specific entropy at that time, for ejected trajectories
from model B090. Nuclides with $A \geq 12$ are counted as seeds. The gray dashed
line indicates the minimum value of $Y_\text{n}/Y_\text{seed} = 35$ that is
required to make lanthanides and actinides. All trajectories with
$Y_{e,\text{5GK}} < 0.25$ (black dots) are above the minimum neutron-to-seed
ratio regardless of initial entropy, while almost all trajectories with
$Y_{e,\text{5GK}} > 0.30$ (blue dots) are below it. The remaining trajectories
($0.25 \leq Y_{e,\text{5GK}} \leq 0.30$, red dots) can produce lanthanides or
actinides if the initial entropy is $\gtrsim 30 - 40\ k_B\ \text{baryon}^{-1}$.}
\label{fig:nseed}
\end{figure}

\Cref{fig:laac_heat} also shows that there is a separate population of particles
that contain significant amounts of lanthanides and actinides for electron
fractions in the range $0.25 \lesssim Y_{e,\text{5GK}} \lesssim 0.3$,
particularly in models B090, BF15, and to a lesser extent in model H010. This
group of trajectories has higher entropies ($s_\text{5GK} \gtrsim 30\
k_B\,\text{baryon}^{-1}$) relative to those which do not make lanthanides or
actinides for $Y_{e,\text{5GK}}\geq 0.25$. We can understand this population by
considering the strong dependence of the lanthanide/actinide abundance on the
neutron-to-seed ratio $Y_\text{n}/Y_\text{seed}$, where $Y_\text{seed} =
\sum_{A\geq 12} Y_A$. In general, a neutron-to-seed ratio
$Y_\text{n}/Y_\text{seed} \sim 40$ is necessary to produce a significant amount
of lanthanides and actinides: as $Y_\text{n}/Y_\text{seed}$ increases from about
35 to about 45, $X_\text{La} + X_\text{Ac}$ increases from $\sim 10^{-5}$ to
$\sim 10^{-2}$. This effect is illustrated in \cref{fig:nseed} for model B090,
where the different particle populations are shown in different colors. The
trajectories in the range $Y_{e,\text{5GK}} = 0.25 - 0.30$ lie on both sides of
the $Y_\text{n}/Y_\text{seed} = 35$ boundary for the range of entropies
encountered in the disk ejecta, with a critical entropy for lanthanide
production  $s_\text{5GK}\sim 30 - 40\ k_B\,\text{baryon}^{-1}$. Outside of this
range of $Y_e$, lanthanide production is insensitive to $s_\text{5GK}$. The
neutron-to-seed ratio increases with increasing entropy because a higher entropy
prefers a larger number of particles, and thus the composition contains more
lighter particles such as free neutrons.

The population of trajectories with $Y_{e,\text{5GK}} = 0.25 - 0.30$ and
$s_\text{5GK} \gtrsim 30\ k_B\,\text{baryon}^{-1}$ is absent in model Hinf (cf.~\cref{fig:laac_heat}. In
order to achieve such entropies at modest electron fraction, neutrino
irradiation cannot be too strong. In model Hinf, the weak interaction timescale
is comparable to the viscous heating timescale, raising $Y_e$ above $0.3$. Since
the \HMNS{} is the strongest source of neutrinos while it persists, this
population of trajectories can only exist if there is either no \HMNS{} or if
the \HMNS{} collapses to a \BH{} quickly. The eight trajectories in model Hinf
that produce lanthanides and actinides at $Y_{e,\text{5GK}} \gtrsim 0.4$ are an
extreme case. They all have $s_\text{5GK} > 200\ k_B\,\text{baryon}^{-1}$, which
allows the neutron-to-seed ratio to be high enough to make lanthanides and
actinides even at $Y_e > 0.4$. These particles attain this high entropy because
of late-time fallback into (and then rapid ejection from) the innermost part of the disk, where they experience
significant heating past $5$~GK.

\begin{figure*}
\includegraphics[width=\linewidth]{%
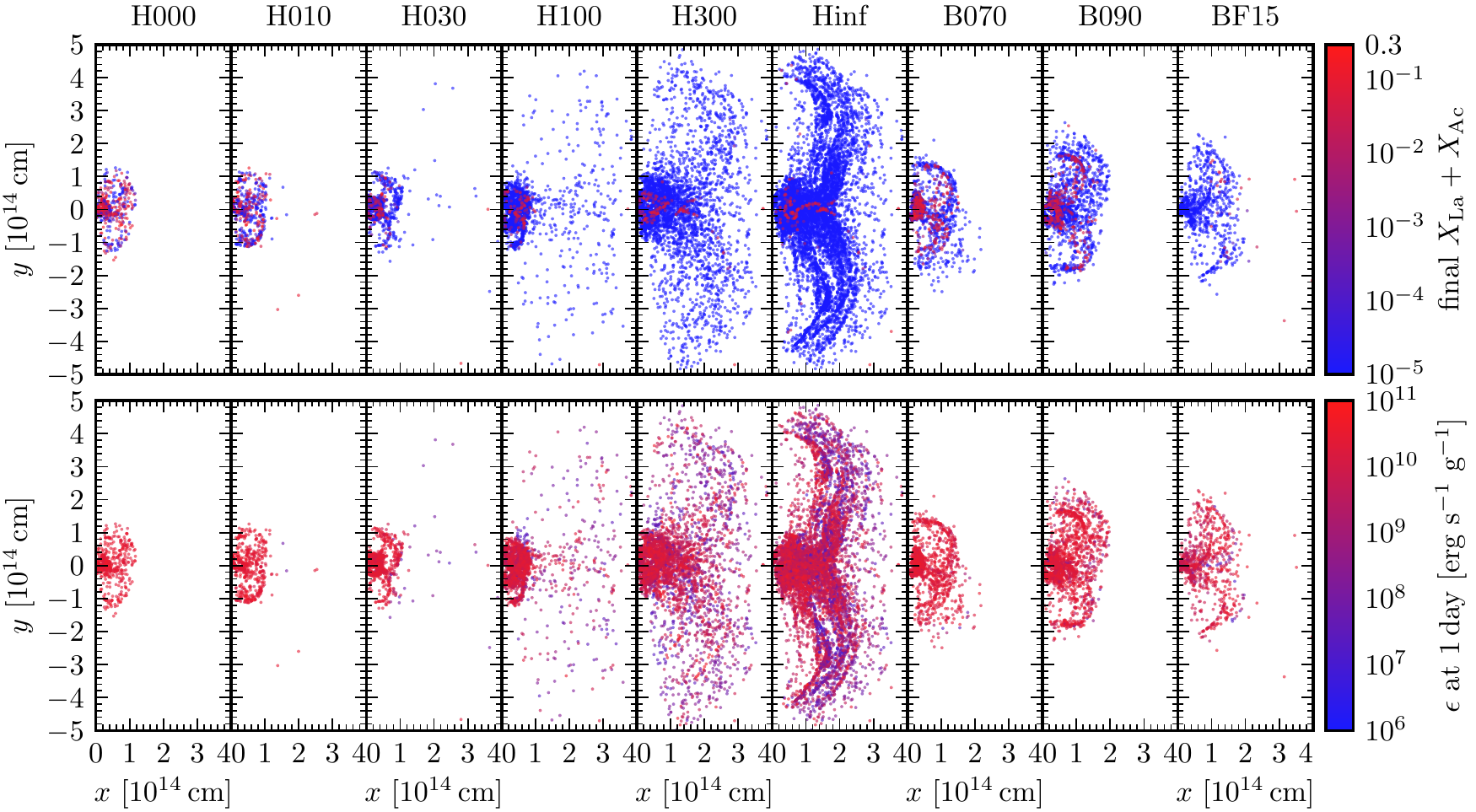}
\caption{Spatial distribution of ejected trajectories at time $t=1$~day, for all
\HMNS{} and \BH{} models with non-zero viscosity. Positions are computed by
assuming that the velocity of each particle at the end of the simulation ($\sim
10$~s) remains constant at later times. Aside from smoothing of sharp features
by radioactive heating at late times \citep{rosswog:14,fernandez:15}, this
extrapolation of trajectories yields a good first approximation to the
morphology of the homologously expanding disk wind ejecta. \emph{Top:} Final
lanthanide and actinide mass fraction. Because the high-lanthanide component of
the ejecta is very small in most models, the points with $X_\text{La} +
X_\text{Ac} \geq 10^{-3}$ are plotted above all other points with lower mass
fractions (except for models H000 and H010). The lanthanide and actinide mass
fractions stop changing a few seconds after neutrons are exhausted. Therefore,
the final mass fractions are the same as the ones prevailing during the kilonova
emission phase. \emph{Bottom:} Radioactive heating rate at one day. In this row,
the points for the different trajectories are drawn in a random order for all
models.}
\label{fig:spatial}
\end{figure*}

\Cref{tab:ejecta} gives the total radioactive heating rate in the ejecta $\epsilon_\text{tot}$ at 1 day
and at 7 days for all models. The quantity scales
with the total ejecta mass. When considering the contribution of the disk
outflow to a radioactively-powered kilonova, we need to keep in mind the
contribution of the dynamical ejecta, which is not simply additive as in the
case of nucleosynthesis. The larger fraction of lanthanides and actinides
expected for the dynamical ejecta means that its optical opacity is likely to be
larger than that of the disk outflow. The geometry of the dynamical ejecta
depends on the type of merger involved: quasi-spherical for NSNS mergers
\citep[e.g.,][]{hotokezaka:13}, or confined to the equatorial plane for NSBH
mergers \citep[e.g.,][]{kawaguchi:15,foucart:17a}. In the former case, the disk
ejecta can be obstructed in all directions by high-opacity material, whereas in
the latter a long-lasting blue optical component from the disk ejecta can be
detectable from some directions \citep{kasen:15}. A short-lived blue
optical component should be detectable in most cases since the lanthanide-rich
material goes through a high-temperature phase and its outer layers let photons
escape more freely \citep{barnes:13,fernandez:17a}

If we only consider the heating rates and lanthanide/actinide content from the
disk outflow, we expect models H000 and H010 to make dim infrared kilonovae,
while all other models should make bright blue optical kilonovae. The exception
is B070, which has about 1.5\% lanthanides and actinides by mass and a fairly
large heating rate, so this model could make a brighter infrared kilonova. The
heating rates reported here are upper limits to the bolometric luminosities of
kilonovae from the disk outflow, since the conversion of energy from
radioactive decay products into thermal energy of the ejecta has a limited
efficiency \citep[e.g.,][]{metzger:10b,hotokezaka:16,barnes:16}.

The bottom row of \cref{fig:laac_heat} shows the heating rates of the individual
trajectories at $t=1$~day as a function of $Y_{e,\text{5GK}}$. The heating rate
decreases slowly with increasing electron fraction for $Y_{e,\text{5GK}}
\lesssim 0.3$. For higher initial $Y_e$, the decrease steepens until large
oscillations start around $Y_{e,\text{5GK}} \sim 0.4$. This general behavior is
consistent with the findings of the parametrized r-process nucleosynthesis study
of \citet{lippuner:15}. The oscillations are due to the initial \NSE{}
composition, which can be dominated by a small number of individual nuclei that
match the electron fraction of the material. If there is a nuclide with a
matching $Y_e$ that has a half-life of about a day, there will be strong heating
at $t\sim 1$~day as this nuclide decays, but if there is no such nuclide, then
there will be significantly less heating. The peak in the heating rate at
$Y_{e,\text{5GK}} \sim 0.425$, visible in all but the first column of
\cref{fig:laac_heat}, is caused by \smce{^{66}Ni}, which has $Y_e = 28/66 \sim
0.424$ and is a dominant nuclide in the initial \NSE{} composition.
\smce{^{66}Ni} decays to \smce{^{66}Cu} with a half-life of 55 hours, which then
decays to stable \smce{^{66}Zn} with a half-life of 5 minutes. Around
$Y_{e,\text{5GK}} \sim 0.45$, there is a large dip in the heating rate because
the initial composition is dominated by \smce{^{62}Ni} and \smce{^{66}Zn}, both
of which are stable and have $Y_e = 0.45$. The little neutron capture that takes
place mainly builds up \smce{^{88}Sr}, which is also stable and has $Y_e =
0.43$. At $Y_{e,\text{5GK}} \sim 0.49$ there is another peak in the heating rate
due to \smce{^{62}Zn}, \smce{^{61}Cu}, and \smce{^{57}Ni}, which are all
unstable and very abundant since they have electron fractions between 0.48 and
0.49. At higher entropies, the initial neutron to seed ratio is enhanced, which
will generally produce unstable nuclei and thus possibly break the oscillatory
pattern of heating rate versus\ $Y_{e,\text{5GK}}$.

The morphology of the ejecta for all models with non-zero viscosity is
shown in \cref{fig:spatial}, with the top row showing final lanthanide and
actinide mass fractions. Since the material is close to homologous expansion,
\cref{fig:spatial} essentially shows the velocity space distribution of the
ejected particles. In all models, most of the particles are concentrated in a
central blob. When the \HMNS{} persists for $100$~ms or more, some particles
attain higher velocities ($v \sim 0.1\,c$) and organize themselves into multiple
shells \citep[which arise from \emph{shock trains} in the gas;
e.g.,][]{matsuo:99}. Since particles are accelerated to high velocities
predominantly by neutrino irradiation, there are more high-velocity particles
the longer the \HMNS{} persists.

In some models (H000, H010, and B070) there appears to be little structure in
the distribution of lanthanides and actinides. In others there is a clear
preference for the high-lanthanide trajectories to be at low velocities and
clustered in the equatorial plane. Model BF15 is an exception in that the few
particles that make a significant amount of lanthanides and actinides achieve
the highest velocities, because they experienced a significant amount of heating
(cf.\ \cref{fig:laac_heat}). The few high-entropy trajectories in model Hinf
that make lanthanides and actinides at $Y_{e,\text{5GK}} \gtrsim 0.4$ are the
high-velocity points moving predominantly in the polar direction, i.e., they
have a large absolute $y$ coordinate and a small $x$ coordinate in
\cref{fig:spatial}. For the heating rates, on the other hand, there is no strong
spatial correlation between heating magnitude and trajectory location in any of
the models. Models H300 and Hinf are the only ones which have a significant
component of low heating rate trajectories, although these particles follow the
same velocity distribution as those with high heating rates. Note that the
low-heating particles are easier to see in the outer part of the ejecta because
the lower particle density, but they are also present in the central blob.

\subsubsection{Comparison with r-process abundances in metal-poor halo stars}

\begin{figure}
\includegraphics[width=\columnwidth]{%
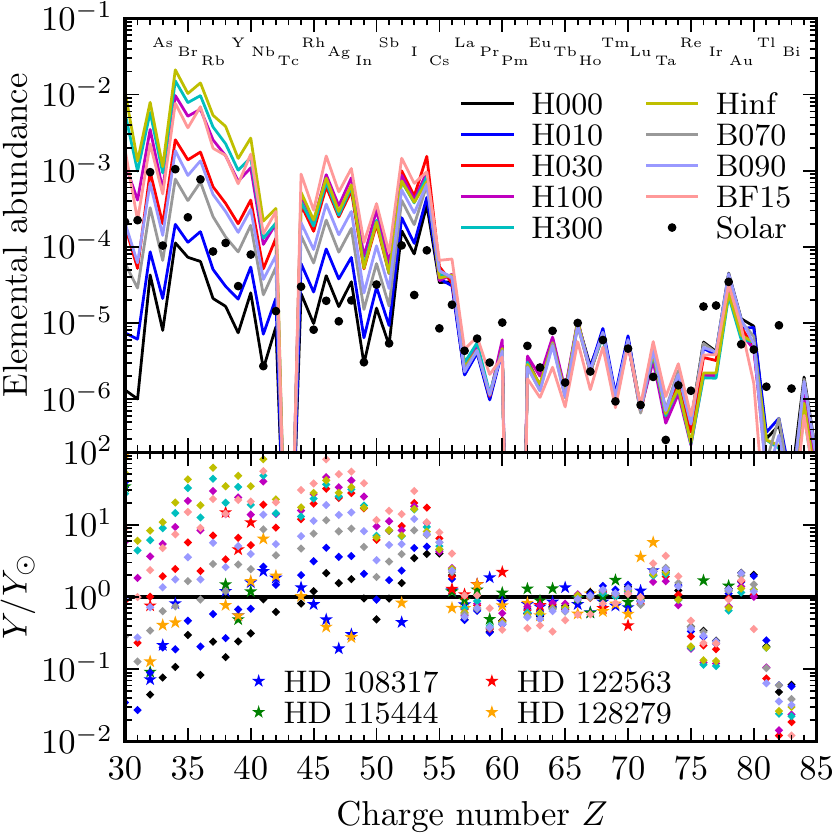}
\caption{\emph{Top:} Final elemental abundances for all models with non-zero
viscosity. In each case, abundances are scaled so that $\sum (\log
Y_Z/Y_{Z,\odot})^2$ is minimized for $55 \leq Z \leq 75$. Black dots show solar
system r-process abundances \citep{arnould:07}. \emph{Bottom:} Ratio of model
abundances to solar system r-process abundances (diamonds). Also shown are the
observed abundances of four metal-poor halo stars \citep[shown as stars, from][]{westin:00,roederer:12}, scaled to
the solar system values in the same way as model abundances.}
\label{fig:z_abund}
\end{figure}

Metal-poor galactic halo stars that show r-process elements in their spectra
are thought to have been enriched by a single or few r-process nucleosynthesis
events \citep{cowan:99}. \Cref{fig:z_abund} shows the final elemental abundances
(at $t \sim 10$~Myr) of our investigated models, along with the solar r-process
abundances from \citet{arnould:07} and from four metal-poor halo stars as
reported in \citet{westin:00} and \citet{roederer:12}. Abundances are scaled to
give the best possible match to the solar values in the range $56 \leq Z \leq
75$. There is little variation between the different models in this range, and a
generally good match to solar and metal-poor star abundances. While Eu ($Z =
63$) is under-produced by a factor of a few with respect to solar in our models,
the metal-poor halo stars also exhibit slightly sub-solar Eu abundances.

\pagebreak[4]
Regarding elements in the range $44 \leq Z \leq 55$, \cref{fig:z_abund} shows
that they are overproduced relative to the solar values by all models except
H000. In the range $Z = 30-42$, our disk outflow models do not match the solar
abundances, but nonetheless agree with the overall increase in abundances with
$Z$ (compared to solar) exhibited by metal-poor halo stars. From $Z=44$ to $47$,
the metal-poor halo stars have a declining abundance trend compared to solar,
which none of our models reproduce. In fact, the disk outflow generates an
increasing trend from $Z = 44$ to 47. Thus, we can conclude that both the solar
and metal-poor halo stars' abundance patterns are inconsistent with pure disk
outflow nucleosynthesis. This means that the nucleosynthesis event (or events)
that enriched these metal-poor halo stars in heavy r-process elements could not
have been the r-process in merger disk outflow alone. There must have been
contributions from additional types of ejecta, or perhaps a different kind of
nucleosynthesis event altogether. Contributions from very neutron-rich dynamical
ejecta that produces mainly second to third peak material ($Z \gtrsim 55$) could
make the combined final abundance pattern consistent with the metal-poor halo
star observations \citep[cf.][]{just:15}. For $Z \lesssim 42$, we expect
supernovae to contribute to the nucleosynthesis, thus making it difficult to
draw any conclusions from the disk outflow nucleosynthesis alone.

\subsection{Impact of angular momentum transport}

\begin{figure}
\includegraphics[width=\columnwidth]{%
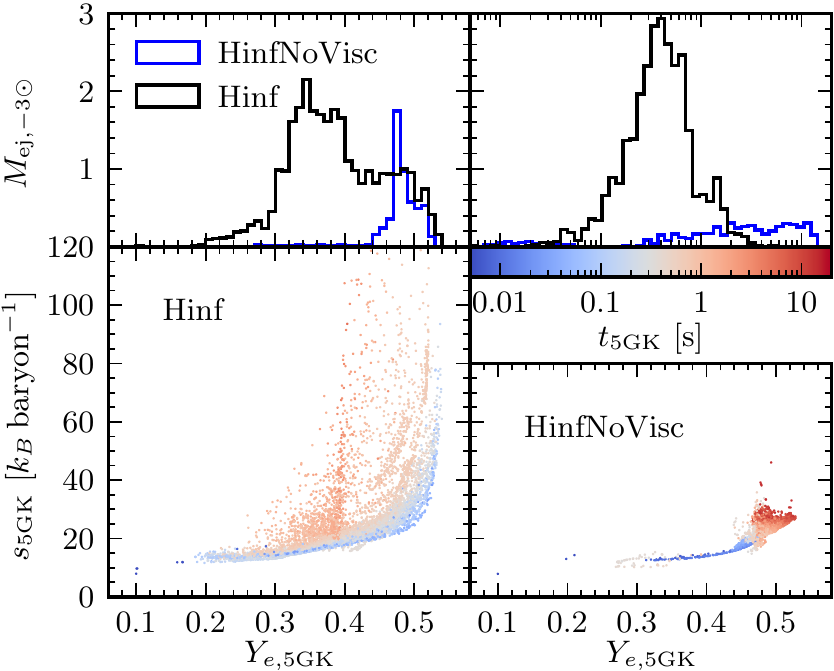}
\caption{Properties of the disk wind ejecta from a \HMNS{} that does not
collapse, including and excluding the effect of viscosity on angular momentum
transport and heating (models Hinf and HinfNoVisc, respectively). \emph{Top:}
Histograms of the ejected mass in units of $10^{-3}\,M_\odot$ as a function of
$Y_{e,\text{5GK}}$ (left) and as a function of the time $t_\text{5GK}$ when each
particle has cooled to 5~GK (right). \emph{Bottom:} Scatter plot of
$Y_{e,\text{5GK}}$ versus \ $s_\text{5GK}$, color-coded by $t_\text{5GK}$ for
model Hinf (left) and HinfNoVisc (right).}
\label{fig:scat_visc}
\end{figure}

Neutrino irradiation from a \HMNS{} is strong enough that significant amounts of
material can in principle be ejected by neutrino energy deposition alone
\citep{ruffert:97}. Therefore, it is useful to clarify the contribution of
angular momentum transport processes to the overall mass ejection from the disk.
Transport of angular momentum modifies the evolution of the disk in two ways:
(1) it causes part of the disk to accrete and part to spread outward as the
local contribution of centrifugal forces to hydrostatic balance is modified, and
(2) it dissipates energy in the form of heat, increasing the local entropy of
the gas. While the detailed form of these two effects depends on
the way the process is modeled (i.e. $\alpha$ viscosity), the overall
contribution to the disk evolution is generic \citep[e.g., magnetohydrodynamic
turbulence also heats up the gas, but with a different spatial distribution than
shear viscosity; e.g.,][]{hirose:06}.

We focus here on two models of a \HMNS{} which does not collapse: one including
angular momentum transport through $\alpha$ viscosity like all other models
(Hinf), and another in which the viscosity is set to zero (HinfNoVisc).
\Cref{fig:scat_visc} shows the distribution of the ejected particles as a
function of initial electron fraction $\ye$, initial specific entropy $s_\text{5GK}$, and
nucleosynthesis start time ($t_{5KG}$).

\pagebreak[4]
A non-zero viscosity increases the total ejecta mass from $5.5\times
10^{-3}\,M_\odot$ (HinfNoVisc) to $29.6\times 10^{-3}\,M_\odot$ (Hinf), which
corresponds to 18\% and 99\% of the initial disk mass, respectively. Not all of
this additional mass is directly ejected by angular momentum transport, however.
The scatter plot of $\ye$ versus\ $s_\text{5GK}$ in \cref{fig:scat_visc} shows a
clear neutrino-driven wind pattern for part of the ejecta from model Hinf (see
also \S\ref{sec:ejecta_properties} and \cref{fig:scatter}). This pattern is
associated primarily with the early ejecta ($t_\text{5GK} \lesssim 0.2$~s), which
covers the range $\ye = 0.2 - 0.54$. Some particles that start nucleosynthesis
later ($t_\text{5GK} \sim 0.5$~s), with electron fractions in the range $\ye =
0.4-0.54$, also exhibit strong correlations between $\ye$ and $s_\text{5GK}$.
This later ejecta could thus also be neutrino-driven, but the different
thermodynamic conditions prevailing at later times alter the exact relationship
between $\ye$ and $s_\text{5GK}$. Ejecta that begins nucleosynthesis at later
times ($t_\text{5GK} \gtrsim 0.8$~s), with $\ye = 0.3-0.4$, shows a much larger
dispersion in $s_\text{5GK}$ for a given $\ye$. This dispersion is associated
with convective motions in the advective phase of the disk, in which mass
ejection is driven exclusively by viscous heating and nuclear recombination
(\S\ref{sec:ejecta_properties}).

In contrast, \cref{fig:scat_visc} shows that the model with zero viscosity
ejects almost exclusively material with $\ye = 0.47 - 0.52$, with ejection
happening predominantly at late times ($t_\text{5GK} \gtrsim 1$~s). All of the
trajectories in model HinfNoVisc exhibit the characteristic neutrino-driven wind
correlation between electron fraction and entropy. Given that the model with
viscosity has more neutrino-driven wind ejecta than model HinfNoVisc, we infer
that angular momentum transport not only adds an additional late-time ejecta
component, but it also enhances the neutrino-driven wind itself. This
enhancement arises from two effects. First, spreading of the outer regions of
the disk due to angular momentum transport moves material into shallower regions
of the gravitational potential, where thermal unbinding requires less energy
injection. Second, viscous heating also acts on the neutrino-driven wind ejecta,
increasing the rate of internal energy gain. This also explains why the
neutrino-driven component is delayed in the case without viscosity: particles
are more tightly bound, requiring energy deposition for a longer time in order
to be ejected.

\begin{figure}
\includegraphics[width=\columnwidth]{%
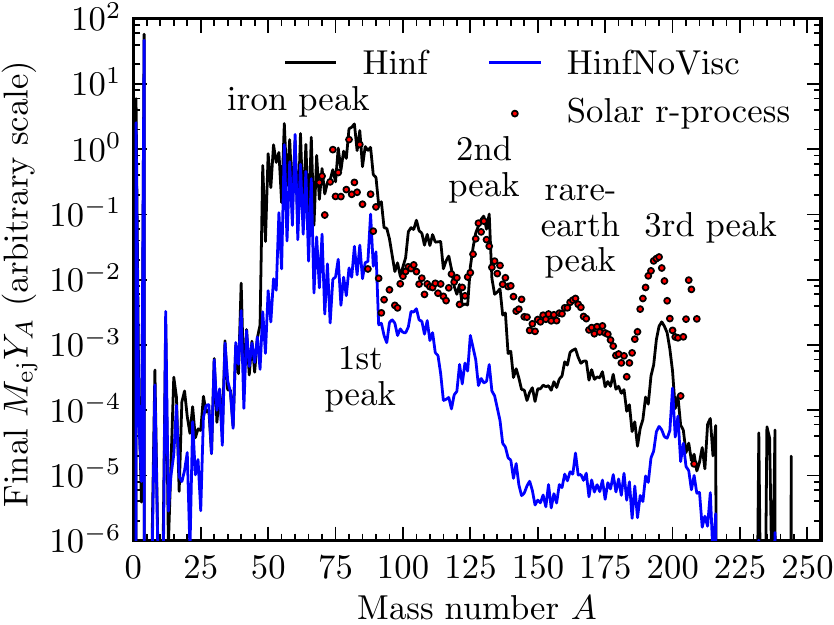}
\caption{Final abundances of non-collapsing \HMNS{} models with and without
viscosity (Hinf and HinfNoVisc, respectively). Model abundances are scaled by
the total ejecta mass of each model, while the solar system r-process abundances
\citep{arnould:07} are scaled to match the second peak of model Hinf.}
\label{fig:abund_visc}
\end{figure}

\Cref{fig:abund_visc} shows the final abundances for models Hinf and HinfNoVisc.
Without viscosity, most of the ejecta has $\ye \sim 0.5$, and so the final
abundance is dominated by the iron peak ($A \sim 56$) and \smce{^4He}. Note that
HinfNoVisc produces the same iron peak and \smce{^4He} abundance as Hinf despite
its total ejecta mass being more than a factor of five lower. For elements
significantly heavier than the iron peak, the final abundance of model
HinfNoVisc is one to two orders of magnitude lower than in model Hinf. While
model HinfNoVisc can still make the third r-process peak, the material is
produced by just a handful of particles that have $\ye \leq 0.25$.

The nucleosynthesis in the disk outflow from a \HMNS{} has also been studied by
\citet{martin:15}, who focused on the neutrino-driven wind without the
contribution from angular momentum transport. In their study, they use the
ejection time of a particle as a proxy for the \HMNS{} lifetime.
\citeauthor{martin:15} find that the total ejecta mass increases with increasing
\HMNS{} ejection time, and that the ejecta also becomes less neutron-rich at
later times. These results are broadly consistent with what we find in model
HinfNoVis.

A more detailed comparison between the results from \citeauthor{martin:15} and
ours shows that the initial disk mass can have a significant impact on the
properties of this ejecta component. \citeauthor{martin:15} find most of the
ejecta having $Y_e \sim 0.3-0.4$, and a majority of the particles ejected on a
timescale of $100$~ms, whereas model HinfNoVisc has $Y_e\sim 0.5$ and most
trajectories begin nucleosynthesis after 1~s. The initial condition in
\citeauthor{martin:15} is the output of a three-dimensional neutron star merger
simulation by \citet{perego:14}, with a disk that is more than six times as
massive as in our model ($0.19\,M_\odot$ compared to $0.03\,M_\odot$). Different
physical and thermodynamic conditions in the disk can lead to
different results in the nucleosynthesis (e.g., compare the output of our models
B070 with BF15). Another source for the differences in the ejecta properties can
be the different hydrodynamic methods used, which can have different amounts of
numerical viscosity. Finally, one similarity between the results of
\citeauthor{martin:15} and ours is the characteristic values for the entropy in
the neutrino-driven wind \citep[$\sim 20\ k_B\ \text{baryon}^{-1}$; compare the
bottom right panel of \cref{fig:scat_visc} and Figure 5 in][]{martin:15}.

\section{Conclusions}
\label{s:conclusions}

We have performed nucleosynthesis calculations in the outflow from a neutron
star merger accretion disk when the central object is a \HMNS{} or a \BH{}.
We used long-term hydrodynamic simulations of accretion disks to model the
ejecta, and detailed nucleosynthesis calculations were carried out on tracer
particles using a nuclear reaction network. We have systematically varied the
lifetime of the central \HMNS{} to study its impact on the nucleosynthesis. Our
simulations continued after the \HMNS{} collapses to a \BH{}, thus allowing us to
investigate the long-term effects on the disk and nucleosynthesis. We have also
performed some simulations that start with central \BH{s} with different spins,
to explore similarities and differences with the \HMNS{} case.

Our results are consistent with previous findings regarding the monotonic
increase in mass ejection and mean electron fraction of the disk outflow for
longer \HMNS{} lifetimes \citep[\cref{fig:ye_hist_hmns}; see
also][]{metzger:14}. This correlation results in the amount of ejecta that has
initial $Y_e \leq 0.25$ being almost constant once the \HMNS{} lifetime is $\tau
\gtrsim 30$~ms. The final abundance pattern at large mass numbers ($A\gtrsim
100$), normalized by the total ejecta mass, is thus independent of the \HMNS{}
lifetime, because only material with $Y_e \lesssim 0.25$ can make those heavy
elements (\cref{fig:abund}). For very short \HMNS{} lifetimes ($\tau \lesssim
10$~ms), there is more neutron-rich ejecta and thus the rare-earth and third
r-process abundance peaks are about half an order of magnitude larger. For these
short lifetimes, the disk outflow abundances alone are almost consistent with
the solar r-process abundances.

For other cases, the inconsistency between the final abundances from the disk
outflow and the solar values is not in itself problematic. In most neutron star
mergers, we also expect a dynamical ejecta component that tends to be more
neutron-rich. This means it could easily synthesize the heavy r-process material that
the disk ejecta may be under-producing. If the dynamical ejecta is consistent
with the solar r-process abundances between the second and third peak, then our
results indicate a preference for short \HMNS{} lifetimes, since abundances from
these models are broadly consistent with solar r-process abundances between the
second and third peak. If, on the other hand, the dynamical ejecta over-produces
the third peak relative to the second peak, then we would require longer \HMNS{}
lifetimes, which result in under-production of the third peak compared to the
second, in order to make the combined abundance pattern consistent with solar.
We draw similar conclusions from comparing the nucleosynthesis in our disk
models to the observed abundances in metal-poor halo stars (\cref{fig:z_abund}).

If the \HMNS{} lifetime is $\tau \geq 30$~ms, most of the ejected material has
$Y_e \gtrsim 0.25$ and so only a small amount of lanthanides and actinides are
synthesized. Thus we expect the disk outflow to produce a kilonova that
peaks in the optical band on a timescale of a day. However, this optical
kilonova component may be obscured by lanthanides and actinides that were
synthesized in the more neutron-rich dynamical ejecta, particularly because this
ejecta component is quasi-spherical in NSNS
mergers \citep[e.g.,][]{bauswein:13}. For a BHNS merger on the other hand,
viewing-angle effects are expected to be very important given that the
dynamical ejecta should lie primarily in the equatorial plane, allowing
emission from lanthanide-free material to more easily escape in the polar
direction \citep[e.g.,][]{kasen:15}. However, BHNS mergers do not make
\HMNS{s}, so visibility of the \HMNS{} component relies on the dynamical ejecta
from the NSNS binary having a small mass.
If the \HMNS{} lifetime is
$10$~ms or less, then the disk outflow contains substantial amounts of
lanthanides and actinides. This would also produce an infrared kilonova on a
timescale of a week and could be indistinguishable from the dynamical
ejecta contribution.
Due to the possible masking of optical emission by the dynamical ejecta,
the absence of optical emission could only be an indication for a short-lived
\HMNS{} if the viewing-angle can be tightly constrained to be face-on, e.g.\
by a coincident \sGRB{} detection.
The properties of the radioactive heating rates from our
tracer particles (\cref{fig:laac_heat}) are consistent with the findings of
\citet{lippuner:15}. We find no strong spatial correlation between radioactive
heating and particle location in the ejecta (\cref{fig:spatial}).

By comparing the disk outflow abundances when a spinning \BH{} or a \HMNS{} is
the central object, we find that a \BH{} of spin $0.7-0.9$ can mimic a \HMNS{}
with lifetime $\tau\sim 15$~ms. If the central \BH{} is more massive (and the
disk is compact, such as in a NSBH remnant), then it can mimic a longer-lived
\HMNS{} with $\tau \sim 60$~ms. For longer \HMNS{} lifetimes ($\tau \gtrsim
100$~ms), we find no possible overlap between the amounts of mass ejected or the
nucleosynthesis signatures. Determining whether this difference translates into
the kilonova signature requires more detailed radiation transport calculations
using realistic initial conditions for the central object, disk, and dynamical
ejecta.

Regarding the disk outflow dynamics, we find that when a \HMNS{} is present, two
types of ejecta are clearly distinguishable: one driven primarily by neutrino
energy deposition, showing a clear correlation between electron fraction and
entropy, and another driven primarily by viscous heating and nuclear
recombination, in which no such correlation exists (\cref{fig:scatter}). We
evolved a test model with no explicit angular momentum transport, and found that
the contribution of viscosity is not limited to the late-time
neutrino-independent outflow. Angular momentum transport also aids the ejection
of the neutrino-driven component by moving material to shallower regions of the
gravitational potential, and by accelerating thermal ejection through additional
heating of the gas (\cref{fig:scat_visc}). Thus, excluding angular momentum
transport in the \HMNS{} disk evolution can severely underestimate the amount of
mass ejected.

While our hydrodynamics simulation includes the most relevant physics
(neutrino transport, viscosity, and general relativity), all of these are
currently implemented in an approximate way.
Neutrino interactions in the disk play a crucial role in setting the
electron fraction and specific entropy distributions of the outflow.
Likewise, the dominance of the \HMNS{} irradiation in driving the
early outflow requires a realistic treatment of the merger remnant
instead of its approximation as a reflecting boundary. Hence our
approximate treatment of the \HMNS{} surface and neutrino interactions
are the most important limitations of this study.
We believe that our approximate neutrino treatment captures the most
important aspects of the neutrino interactions in the disk and so we expect that
our results exhibit the correct general trends. However, due to the importance
of neutrinos in driving the disk outflow, it is clear that more sophisticated
neutrino radiation transport methods are needed in future work.
Incorporating an $\alpha$ viscosity
prescription into our disk simulations is a significant step forward
from simulations with no physical viscosity. Ultimately, however,
accretion disk simulations with more realistic physical viscosity
(e.g., provided by magnetoturbulence driven by the magnetorotational
instability) need to be performed. Furthermore, due to the extreme
computational complexity of fully general-relativistic hydrodynamics
simulations, no self-consistent simulations of neutron star mergers
with subsequent long-term accretion disk evolution have been carried
out to date. Our approach of assuming an initial equilibrium torus is
thus a necessary approximation in order to investigate nucleosynthesis
in merger accretion disks. Mapping the early accretion disk structure
found in merger simulations into another code for the disk evolution
(as we have done for model BF15) is a more accurate approach, but it
is still not fully self-consistent. With next generation
general-relativistic hydrodynamics codes, we hope to be able to
perform completely self-consistent simulations of neutron star mergers
and accretion disk outflows in the future.  Finally, we only use the
FRDM nuclear mass model for the nucleosynthesis calculations in this
study. In future work, we plan to investigate and compare other
nuclear mass models for r-process nucleosynthesis in disk outflows.

\section*{Acknowledgments}

JL and CDO acknowledge support from NSF grant AST-1333520.
RF acknowledges support from the University of California Office of
the President, from NSF grant AST-1206097, and from the Faculty of
Science at the University of Alberta.
DK is supported in part by a Department of Energy Office of Nuclear
Physics Early Career Award, and by the Director, Office of Energy
Research, Office of High Energy and Nuclear Physics, Divisions of
Nuclear Physics, of the U.S. Department of Energy under Contract No.\
DE-AC02-05CH11231.
The software used in this work
was in part developed by the DOE NNSA-ASC OASCR Flash Center at the
University of Chicago. This research used resources of the National
Energy Research Scientific Computing Center (NERSC), which is
supported by the Office of Science of the U.S.\ Department of Energy
under Contract No. DE-AC02-05CH11231. Some computations were performed
on the \emph{Edison} compute cluster (repository m2058).  Some of the
calculations were performed on the \emph{Zwicky} compute cluster at
Caltech, supported by NSF under MRI award PHY-0960291 and by the
Sherman Fairchild Foundation. This work was supported in part by NSF
grant PHY-1430152 (JINA Center for the Evolution of the Elements).
CDO thanks the Yukawa Institute for Theoretical Physics for support
and hospitality. This article has been assigned Yukawa Institute report
number YITP-17-26.

\appendix

\section{Improvements to the hydrodynamic disk simulations}

We have corrected an error in the weak interaction rates
used since \citet{fernandez:13}. The error involved the absence
of the neutron-proton mass difference in the argument of the Fermi-Dirac
integrals for all the electron neutrino rates (but not in the antineutrino rates).
After correction of this error, the electron neutrino emission rates
decrease by a factor of up to two in regions with temperatures $\gtrsim 1$~MeV.
The most significant change in the result is an increase in the fraction
of matter ejected with $Y_e > 0.25$ for the non-spinning \BH{} case from a few percent to
$\sim 30\%$ of the outflow, resulting in an increase of $\sim 10\%$ in
the average $Y_e$ of the outflow. A similar increase in the mean $Y_e$
of the wind is observed for the long-lived \HMNS{} case.
In terms of the amount of mass ejected,
the effects are strongest for a non-spinning \BH{}, which ejects up to
$\sim 10\%$ more mass than before the correction is applied (i.e. an additional
$\sim 0.5\%$ of the initial disk mass). For a long-lived \HMNS{}, the
mass ejection increases by $\sim 0.2\%$.

Relative to \citet{metzger:14}, we also account for separate neutrino and
antineutrino temperatures for absorption. The mean energy for each neutrino species
is calculated by taking the ratio of the globally-integrated
energy to number emission rates, as in \citet{ruffert:96}.
The temperatures are then obtained through the relation
$kT_{\nu,i} = F_4(0)/F_5(0) \langle \varepsilon_{\nu,i}\rangle$,
where $F_i$ are the Fermi-Dirac functions of integer argument, $F_5(0)/F_4(0) \simeq 5.065$,
and $\langle \varepsilon_{\nu,i}\rangle$ is the neutrino mean energy.
When using the updated weak interaction tables, this change translates into
an \emph{increase} of $\sim 10\%$ in mass ejection for a non-spinning \BH{}, and
a \emph{decrease} of $0.4\%$ in mass ejection for a long-lived \HMNS{}.

To put the effect of these modifications in perspective for the non-spinning \BH{} case,
doubling the resolution of the simulations in each dimension can change mass ejection
by $\sim 10\%$ \citep{fernandez:15}, and
a change in the $\alpha$ viscosity parameter leads
to an almost directly proportional change in mass ejection \citep[i.e., a factor ten when
going from $\alpha = 0.01$ to $\alpha=0.1$,][]{wu:16}.


\bibliographystyle{hmns}
{\raggedright
\bibliography{%
bibliography/nucleosynthesis_references,%
bibliography/eos_references,%
bibliography/grb_references,%
bibliography/ns_references,%
bibliography/gw_detector_references,%
bibliography/NSNS_NSBH_references,%
bibliography/methods_references,%
bibliography/nu_interactions_references,%
bibliography/bh_formation_references,%
bibliography/sph_references,%
bibliography/pns_cooling_references,%
bibliography/sn_theory_references,%
bibliography/populations_references,%
bibliography/accretion_disk_references.bib,%
bibliography/numrel_references.bib,%
bibliography/fluid_dynamics_references.bib,%
bibliography/mhd_references.bib}
}

\end{document}